\documentclass[twocolumn]{aastex631}

\newcommand{\ltsima} {$\; \buildrel < \over \sim \;$}
\newcommand{\gtsima} {$\; \buildrel > \over \sim \;$}
\newcommand{\lta} {\lower.5ex\hbox{\ltsima}}
\newcommand{\kmsMp}{km s$^{-1}$\,Mpc$^{-1}$}
\newcommand{\gta} {\lower.5ex\hbox{\gtsima}}

\newcommand{\kms}{km\,s$^{-1}$}

\newcommand{\cip}{[C\,{\sc i}]~$^{3}$P$_{1}$-$^{3}$P$_{0}$}
\newcommand{\ci}{[C\,{\sc i}]}
\newcommand{\cii}{[C\,{\sc ii}]}
\newcommand{\civ}{[C\,{\sc iv}]}

\begin{document}

\title{The SUPERCOLD-CGM survey: \\I. Probing the extended CO(4–3) Emission of the Circumglactic medium in a sample of 10 Enormous Ly$\alpha$ Nebulae at $z\sim2$.}

\correspondingauthor{Zheng Cai}
\email{zcai@mail.tsinghua.edu.cn}

\author{Jianrui Li}
\affil{Department of Astronomy, Tsinghua University, Beijing, China, 100084}

\author{Bjorn\,H.\,C. Emonts}
\affil{National Radio Astronomy Observatory, 520 Edgemont Road, Charlottesville, VA 22903}

\author{Zheng Cai}
\affil{Department of Astronomy, Tsinghua University, Beijing, China, 100084}

\author{Jianan Li}
\affil{Department of Astronomy, Tsinghua University, Beijing, China, 100084}

\author{Fabrizio Arrigoni Battaia}
\affil{Max-Planck-Institut f\"ur Astrophysik, Karl-Schwarzschild-Str 1, D-85748 Garching bei M\"unchen, Germany}

\author{Jason X Prochaska}
\affiliation{University of California, 1156 High Street, Santa Cruz, CA 95064, USA}
\affiliation{Kavli Institute for the Physics and Mathematics of the Universe,
The University of Tokyo, 5-1-5 Kashiwanoha, Kashiwa, 277-8583, Japan}
\affiliation{Division of Science, National Astronomical Observatory of Japan,
2-21-1 Osawa, Mitaka, Tokyo 181-8588, Japan}

\author{Ilsang Yoon}
\affil{National Radio Astronomy Observatory, 520 Edgemont Road, Charlottesville, VA 22903}

\author{Matthew\,D. Lehnert}
\affil{Universit\'{e} Lyon 1, ENS de Lyon, CNRS UMR5574, Centre de Recherche Astrophysique de Lyon, F-69230 Saint-Genis-Laval, France}

\author{Craig Sarazin} 
\affil{Department of Astronomy, University of Virginia, 530 McCormick Road, Charlottesville, VA 22904, USA}

\author{Yunjing Wu}
\affil{Department of Astronomy, Tsinghua University, Beijing, China, 100084}

\author{Mark Lacy}
\affil{National Radio Astronomy Observatory, 520 Edgemont Road, Charlottesville, VA 22903}

\author{Brian Mason}
\affil{National Radio Astronomy Observatory, 520 Edgemont Road, Charlottesville, VA 22903}

\author{Kyle Massingill} 
\affil{School of Earth and Space Exploration, Arizona State University, Tempe, AZ 85287, USA}

\begin{abstract}

To understand how massive galaxies at high-$z$ co-evolve with enormous reservoirs of halo gas, it is essential to study the coldest phase of the circum-galactic medium (CGM), which directly relates to stellar growth. The SUPERCOLD-CGM survey is the first statistical survey of cold molecular gas on CGM scales. We present ALMA+ACA observations of CO(4-3) and continuum emission from 10 Enormous Ly$\alpha$ Nebula (ELANe) around ultraluminous type-I QSOs at $z\sim2$. We detect CO(4-3) in 100$\%$ of our targets, with 60$\%$ showing extended CO on scales of 15$-$100 kpc. Q1228+3128 reveals the most extended CO(4-3) reservoir of $\sim$100 kpc and is the only radio-loud target in our sample. The CO reservoir is located along the radio axis, which could indicate a link between the inner radio-jet and cold halo gas. For the other five radio-quiet ELANe, four of them show extended CO(4-3) predominantly in the direction of their companions. These extended CO(4-3) reservoirs identify enrichment of the CGM, and may potentially contribute to widespread star formation. However, there is no evidence from CO(4-3) for diffuse molecular gas spread across the full extent of the Ly$\alpha$ nebulae. One target in our sample (Q0107) shows significant evidence for a massive CO disk associated with the QSO. Moreover, 70$\%$ of our QSO fields contain at least one CO companion, two of which reveal extended CO emission outside the ELANe. Our results provide insight into roles of both the cold CGM and companions in driving the early evolution of massive galaxies.

\end{abstract}

\section{Introduction}
\label{sec:intro}

In the Universe, $\sim$ 90\% of the baryons are found to lie in gas reservoirs outside galaxies, which have an important influence on galaxy formation and evolution \citep{2018Nicastro}. At high redshifts, observations and simulations of the multi-phase gaseous halos surrounding massive galaxies indicate that the circum-galactic medium (CGM) not only contributes to star formation, but also can be enriched by galaxy mergers and feedback processes from the host galaxies, such as outflows driven by active galactic nuclear (AGN) activity \citep{2017Tumlinson}. 

Among the multi-phase CGM, Ly$\alpha$ emitting gas is commonly detected ($T\sim$10$^{4}$ K, n\,$\sim$\, 1-100 cm$^{-3}$; \citealt{Cantalupo2017}). As of last century, the scale of the giant Ly$\alpha$ nebulae around high-redshift radio galaxies was gradually found to stretch $>$100 kpc \cite[e.g.,][]{Djorgovski1987, McCarthy1990, Heckman1991, Eales1993, Pentericci2001, Reuland2003, Villar_Martin2003, Miley2006}. During the past few years, giant Ly$\alpha$ nebulae were detected on even larger scales ($\sim$ 200$-$500 kpc) associated with several quasi-stellar objects (QSOs) \cite[e.g.,][]{Cantalupo2014, Martin2014, Hennawi2015, Cai2017b, Cai2018, Borisova2016, Arrigoni_Battaia2019}. The largest of these halos are generally called Enormous Ly$\alpha$ Nebulae (ELANe), and often trace the overdense environments of massive galaxies and proto-clusters \citep{Cai2016, Cai2017b, Arrigoni_Battaia2018, Nowotka2022}. A crucial question arising is how the central massive galaxies co-evolve with the enormous reservoirs of halo gas. To answer this question, it is essential to study the direct connection between the multi-phase CGM and the stellar growth of massive galaxies by mapping the cold molecular gas ($T \sim$ 10–100 K; n\, $\ga$ 10$^{3}$ cm$^{-3}$), which can potentially fuel widespread star formation within the fields of ELANe.

There is growing evidence from cold gas tracers like the rotational transitions of carbon monoxide CO($J,J-1$), the fine-structure lines of atomic carbon \ci\ , and singly ionized carbon \cii\ \citep{Clark2019} that cold gas can extend tens of kpc into the gaseous halos of massive high--z galaxies. While \cii\ can trace both molecular gas and the cold phase of the diffuse neutral or ionized medium, the ground transitions of CO and [CI] have a critical density of n$_{\rm crit}$\,$\sim$\,500 cm$^{-3}$ and temperature of T\,$\sim$10-100\,K \citep[e.g.,][]{Papadopoulos2004}, and therefore trace cold molecular gas. This allow studying the multi-phase CGM in Ly$\alpha$ nebulae across 3$-$4 orders of magnitude in temperature and density. 
However, it has been difficult to establish how general $-$and in some cases also how robust$-$ the detections of extended cold gas reservoirs are, or how well the different phases of the CGM are mixed. \citet{Cicone2015} found a large reservoir of cold gas across the halo of a luminous quasar at $z$\,=\,6.4, which was initially observed to stretch out to a radius of $\sim$30 kpc from the QSO, but later revised by follow-up observations to a radius of $\sim$10 kpc \citep{Meyer2022}. Perhaps the best studied example of cold molecular halo gas is the Spiderweb radio galaxy at $z$\,$\sim$\,2 \citep{Miley2006}, where a widespread ($\sim$70 kpc) reservoir of cold molecular gas was detected in CO(1-0) \citep{Emonts2016}. Subsequent Atacama Large Millimeter/submillimeter Array (ALMA) observations of \ci\ and CO(4-3) showed that the cold molecular CGM has a carbon abundance and excitation conditions resembling the interstellar medium (ISM) in star-forming galaxies \citep{Emonts2018}, while deep Hubble Space Telescope (HST) imaging revealed diffuse blue starlight on the same $\sim$100 kpc scale \citep{Hatch2008}, indicating that the cold molecular CGM fuels in-situ star formation throughout the halo \citep{Emonts2016}. However, the Spiderweb Galaxy is one of the most massive proto-cluster galaxies in formation, hence it is not clear how representative these results are for the general population of high-$z$ massive galaxies. Nevertheless, evidence for molecular halo gas was found in various other high-redshift radio galaxies \citep{DeBreuck2003,Klamer2004,Emonts2014,Emonts2015,Falkendal2021,Li2021,DeBreuck2022}. Claims of extended molecular gas have been made for a number of other high-$z$ QSOs and massive star-forming galaxies \citep[e.g.,][]{Ginolfi2017,Dannerbauer2017,Frayer2018,Fujimoto2019,F2020,Herrera2021,Akins2022}. In the most extreme case, the ALMA Compact Array (ACA) discovered a molecular gas reservoir across $\sim$200 kpc around a quasar at $z$\,$\sim$\,2.2 \citep{Cicone2021}. These studies stress the importance of short-baseline observations for detecting molecular gas on large scales outside of galaxies.

However, a similar number of studies have been reported in the literature where little to no extended cold gas was detected in the CGM around quasar host galaxies at high-$z$ \cite[e.g.,][]{Coppin2008, Polletta2011, Riechers2011, Novak2020, Fogasy2020, Fogasy2022, Decarli2018, Bechtel20, Decarli2021, Decarli2022, Chen2021, Q2021}. \citet{Decarli2018} presented a survey of \cii\ line emission in a sample of 27 $z>6$ quasars, and found 23 out of 27 targets are unresolved, the other four targets only revealed extended emission on a few kpc scale, which does not show any signs of cold halo structures. Additionally, \citet{Novak2020} found \cii\ emission around high-$z$ quasar host galaxies at $z$\,$\ge$\,6 out to $\sim$10 kpc, but ascribed this to the ISM predicted by hydrodynamical simulations, with also no evidence for cold gas halos. One interesting case is the study of the enormous Ly$\alpha$ nebula of Mammoth-I, where CO(1-0) observations obtained with the Karl G. Jansky Very Large Array (VLA) showed that roughly 50$\%$ of the molecular gas was spread across the CGM \citep{Emonts2019}, while observations of the higher transition line of CO(3-2) detected with the Northern Extended Millimeter Array (NOEMA) suggested that all of the molecular gas is locked up within the ISM of the galaxies in the nebula \citep{Q2021}. This discrepancy was confirmed with simultaneous \ci\ $^{3}P_{1}$-$^{3}P_{0}$ and CO(4-3) data from ALMA \citep{Bechtel20}, reinforcing the idea that the different transitions of CO($J$,$J-1$) may not be equally sensitive for tracing extended molecular gas, especially when the excitation of the gas is low (see also \citealt{Papadopoulos2004}).

Therefore, it remains somewhat uncertain to what extent cold molecular gas is present in the circum-galactic environments of high-$z$ massive galaxies, and how this cold gas can best be observed.

\subsection{The SUPERCOLD-CGM Survey}

To study in a systematic way the role that the cold CGM plays in the evolution of active massive galaxies in the Early Universe, we started a survey entitled SUPERCOLD-CGM: “Survey of Protocluster ELANe Revealing CO/CI in the Ly$\alpha$ Detected CGM”. The SUPERCOLD-CGM survey aims at using the short baseline configurations of ALMA/ACA and VLA to search for extended, low-surface-brightness emission from cold molecular gas surrounding the massive host galaxies of AGN at high-redshift. We do this by targeting simultaneously CO(4-3) and [CI] $^{3}$P$_{1}$\,$-$\,$^{3}$P$_{0}$ with ALMA/ACA, and CO(1-0) with the VLA. One of the main goals of the SUPERCOLD-CGM survey is to study how general the presence of cold molecular gas in the CGM is, and what the true extent of such molecular gas reservoirs is. Reported cases range from a few tens of kpc to hundreds of kpc, but in nearly all cases the signal-to-noise (S/N) is low, and detectability critically depends on the brightness sensitivity provided by shortest baselines in the array \citep{Emonts2016, Li2021, Cicone2021}. Another goal is use the line ratios (r$_{\rm CI/CO(4-3)}$, r$_{\rm CI/CO(1-0)}$ and r$_{\rm CO(4-3)/CO(1-0)}$) to gain an insight into the excitation properties of the cold gas in various regions. This will help us understand whether any cold molecular CGM is diffuse, or similar to the clumpy ISM in galaxies \citep[see][]{Emonts2018}. And finally, by comparing the distribution of CO(1-0), CO(4-3), and \ci\ with the Ly$\alpha$ and \civ\ emission from available integral-field spectroscopy data \citep{Cai2018, Cai2019}, we will investigate whether the cold and warm gas trace the same ISM/CGM, or are decoupled gas reservoirs. This may tell us more about the nature of the cold molecular CGM and the physical processes involved, such as outflows, gas cooling, or tidal galaxy interactions.

As a pilot study, we recently reported a massive molecular outflow and $\sim$100 kpc extended CO(4-3) reservoir in one of our sample ELANe, namely the radio-loud QSO Q1228+3128 at $z\sim2.2$ \citep{Li2021}. The outflow and extended CO reservoir are spatially aligned with an inner (sub-kpc) radio jet \citep{Helmboldt2007}, thus we argued that the large scale CO reservoir is formed when the propagating radio jet drives the outflow, and also enriches or shocks and cools pre-existing dusty halo gas \citep{Li2021}. In the current paper, using ALMA and ACA, we present the first statistical study of cold molecular gas in the full sample of 10 enormous Ly$\alpha$ nebulae around ultraluminous Type-I QSOs at $z = 2.1-2.3$, through observations of CO(4-3) line and continuum emission. The paper is organised as follows. In §2, we introduce our observations and data reduction. In §3, we provide the results of the ALMA observations, focusing on CO(4-3) and continuum emission. In §4, we provide discussions for the extended cold gas, a massive CO rotating disk, and the quasar-companion systems. Finally, our conclusions are summarised in §5. Throughout this paper, when measuring distances, we assume a ${\Lambda}$CDM cosmology with H$_{0} = 70$\,\kmsMp, $\Omega_\textrm{M} = 0.30$ and $\Omega_{\Lambda} = 0.70$.

\section{Observations and Data Reduction}
\label{sec:data}
\subsection{Quasar sample}
\label{ssec:quasars}

The sample presented in this paper consists of ultraluminous type-I quasars at $z\sim$2 that are surrounded by ELANe, which have been mapped by the Keck Cosmic Web Imager (KCWI) \citep{Cai2018, Cai2019}. Out of the 17 observed KCWI targets, we chose 10 nebulae that extend on a spatial scale $>$120 kpc for ALMA/ACA observations (see Table.1). These QSOs were selected from the Sloan Digital Sky Survey (SDSS)-IV/eBOSS database (e.g., Pâris et al. 2018), restricted to 2.1$<z<$2.3 \citep{Cai2018, Cai2019}. This is the largest uniform QSO sample covered with Keck, and the only sample at $z<$2.4, where H$\alpha$ can be observed to further constrain the physical properties of the CGM from the ground \citep{Leibler2018, Cantalupo2014}. 

\subsection{ALMA/ACA observations and data reduction}
\label{ssec:obs}

Our observations were carried out during ALMA Cycle 7, with the observing program ID: 2019.1.01251.S (PI: B. Emonts). The data were taken in Band 4 (125-163 GHz) between 2019 October 1 and 2021 July 7. A combination of a compact configuration of the 12m array with a baseline length range of 15 $-$ 500m and the 7m ACA array was used to optimize surface-brightness sensitivity for detecting emission on scales of about $2^{\prime\prime}-8^{\prime\prime}$. In this paper, we consider emission that is both spatially and kinematically unresolved at the resolution of the 12m data, i.e., $<2^{\prime\prime}$ ($<$15 kpc), as part of the ISM of the QSO and companion galaxies, while emission found on scales of $\ga$2$^{\prime\prime}$ ($\ga$15 kpc) belongs to the CGM. Future deep optical imaging may refine this definition.

For the 12m array, most observations were taken in the C43-2 array configuration with baseline lengths ranging from 15 $-$ 314m, but several observations were taken in a more extended or compact configuration, with the shortest baseline always 15m and a maximum baseline length between 314 and 500m. Most targets were observed for two runs with 15 $-$ 48 minutes on-source per run, except for Q1228+3128 (hereafter Q1228, same for the other sources), Q1227, and Q1230, which were observed for three runs, and Q1416 which was observed for four runs. The number of 12m antennas used in the observations ranged from 43 to 51.

For the 7m ACA array, observations were taken with baselines ranging from 9 to 45 m. Most targets were observed in 5$-$6 scheduling blocks with 37$-$50 minutes on-source per block, except for, Q1228, Q1227, and Q1416 which were observed in 10$-$11 scheduling blocks, and Q1230 which was observed in 8 scheduling blocks. The number of 7m ACA array antennas used ranged from 10 to 12.

The observations were centered at the optical coordinates of each quasar. The spectral setup consisted of four spectral windows (SPWs) with a 1.875 GHz bandwidth per SPW. Two adjacent spectral windows were centered on CO(4-3) ($\nu_{\rm rest}$= 461.04 GHz), given the quasar systemic redshifts reported in Table 1. Another two spectral windows were added to cover the continuum, but also include the \cip\ ($\nu_{\rm rest}$= 492.16 GHz) line. While the \cip\ line was detected for many of our targets, this line is much fainter than CO(4-3). Due to small uncertainties in the redshifts combined with restrictions in tuning the ALMA receivers, we therefore chose to optimize the CO(4-3) detectability, resulting in the \cip\ line falling at the very edge of the receiver band for some targets. A careful analysis of the \cip\ line is therefore left for a next paper.

The data were reduced following the standard reduction steps using the Common Astronomy Software Applications package (CASA; \citealt{CASA2022}). The calibration of the data was performed using the ALMA calibration pipeline 5.6.1-8 \citep{Masters2020}, together with the calibration scripts that were supplied with the archival data by the ALMA observatory. The pipeline calibration was used to create the MeasurementSet (MS) for each observing run, which were then combined into a single MS for each source. The continuum emission of the sources in our sample is too weak to perform self-calibration; we previously attempted this on the strongest continuum source Q1228, but the results did not improve (see \citealt{Li2021}). Therefore we present in this paper all the data products without self calibration applied. The calibrated MSs were subsequently imaged using the CASA task TCLEAN by adopting a natural weighting scheme of the visibilities, which maximizes the surface brightness sensitivity. For deconvolution, we used interactive tclean, which means we selected regions around the QSOs or companions in the interactive tclean Graphical User Interface (GUI), and cleaned the signal in each channel until no significant residuals remained visible in the GUI. For each quasar, we created several datasets using the following procedure:

1. Using TCLEAN, for each quasar field, we first used the dataset obtained from the 12m-array observations (hereafter `12m dataset') with no continuum subtracted to create the dirty 12m datacubes with a channel width of 30 km s$^{-1}$ from the SPWs containing the line emission. These datacubes containing the CO(4-3) emission line were subsequently used to extract the spectrum of the quasar host  galaxies, companion galaxies and possible extended emisssion. We then determined the line-free channels of each 12m data cube, which were used to do the continuum subtraction for both 12m and 7m datasets. 

2. We subtracted the continuum in ($u$,$v$)-space for each quasar using the task UVCONTSUB with the polynomial order of the fit set to FITORDER=1 or 0 for each 12m and 7m ACA dataset. 

3.  Using TCLEAN, for each quasar field, on the one hand we used only the 12m continuum-subtracted datasets to create the datacube with full spatial resolution for CO(4-3) emission. On the other hand, in search for more extended emission at a higher sensitivity, we also imaged the combined continuum-subtracted 12m+7m datasets using natural weighting, while also setting the uvtaper parameter in tclean to apply a circular Gaussian taper with a full width at half the maximum intensity (FWHM) of 18.5 k$\lambda$. This uvtaper further increases the weights of the shorter baselines, which decreases spatial resolution, but increases the surface brightness sensitivity for detecting extended emission. The combination of adding ACA data and using a uvtaper provides the best surface-brightness sensitivity for detecting molecular gas spread on scales of many tens of kpc. The channel width of both the untapered 12m and tapered 12+7m datacubes is 60 km\,s$^{-1}$. To obtain position-velocity (P-V) maps, we binned these channels to best visualize the signal, resulting in a channel width of 180 \kms\ for Q0052, Q1228, Q1416, and to 240 \kms\ for Q2123, while we kept the 60 \kms\ channel width for Q0050, Q0107 and Q2121. Specially, we applied a Hanning smooth to the P-V map of Q0050.

4. Using TCLEAN, we created the multi-frequency synthesis images of the continuum for each quasar field by imaging all the line-free channels of the four SPWs of the 12m dataset before continuum subtraction.

A summary of the observations and the results of these datacubes is given in Table\,\ref{tab:cube}.

\begin{table*}
\caption{ALMA observations and imaging summary.} 
\centering

\movetableright=-1.05in
  \label{tab:cube}

 \scalebox{0.7}{
\begin{tabular}{cccccccccc}
\hline
\hline
Target ID & R.A. &Decl. & Redshift&Redshift& $N_\mathrm{ant}$(12m/ACA) & Beam size (PA)& $\sigma_\mathrm{cube}^{\rm (c)}$& $\sigma_\mathrm{continuum}$[12m]$^{\rm (d)}$ &$i$ mag$^{\rm (b)}$ \\
(SDSS)&(J2000)&(J2000)&CO(4-3)&Ly$\alpha$ $^{\rm (a)}$ &  & [arcsec$\times$arcsec], [deg] & mJy\,beam$^{-1}$ channel$^{-1}$ & mJy\,beam$^{-1}$& \\
\hline
Q0050+0051 & 00:50:21.22 &  +00:51:35.0  & 2.2432 $\pm$ 0.00002&2.241& 46/11  & 2.4$\times$2.0 (69.9) &0.13&0.014&17.82 $\pm$ 0.01\\
Q0050(taper) &-&-&-&-&-& 5.6$\times$5.4 (87.6) &0.23&-&-\\

Q0052+0140 &00:52:33.67  & +01:40:40.8& 2.3101 $\pm$ 0.00003 &2.309& 46/10  & 2.4$\times$2.0 (-72.4) & 0.13&0.013&17.34 $\pm$ 0.01\\
Q0052(taper)&-&-&-&-&-&5.6$\times$5.3 (-1.3)&0.22&-&-\\

Q0101+0201&01:01:16.54 &+02:01:57.4&2.4503 $\pm$ 0.000003&2.451& 44/11   &2.7$\times$2.0 (67.3) &0.13 &0.012&18.10 $\pm$ 0.01\\

Q0101(taper)&-&-&-&-&-&5.8$\times$5.3 (70.0)&0.23&-&-\\

Q0107+0314 &01:07:36.90 & +03:14:59.2&2.2825 $\pm$ 0.00003 &2.280& 45/11   & 2.6$\times$2.0 (-68.8) &0.13 &0.013&18.01 $\pm$ 0.01\\

Q0107(taper)&-&-&-&-&-&5.8$\times$5.3 (-71.0)&0.22&-&-\\

Q1227+2848&12:27:27.48  &+28:48:47.9 &2.2653 $\pm$ 0.00004 &2.265& 43/11  &2.9$\times$2.1 (10.7) &0.13&0.013&17.73 $\pm$ 0.01 \\

Q1227(taper)&-&-&-&-&-&5.9$\times$5.4 (27.1)&0.21&-&-\\

Q1228+3128 &12:28:24.97 & +31:28:37.7 & 2.2218 $\pm$ 0.00006 &2.214 & 51/12 &2.8$\times$2.0 (13.6)&0.17&0.043&15.68 $\pm$ 0.01\\
Q1228(taper)&-&-&-&-&-&8.0$\times$7.5 (24.2) &0.32&-&-\\

Q1230+3320&12:30:35.47 &+33:20:00.5  & 2.3287 $\pm$ 0.00002&2.323& 43/11  &  3.2$\times$2.2 (1.1) &0.15 &0.016&17.76 $\pm$ 0.01\\
Q1230(taper)&-&-&-&-&-&  5.9$\times$5.3 (16.2) &0.25 &-&-\\

Q1416+2649& 14:16:17.38 &+26:49:06.2  & 2.2990 $\pm$ 0.00007&2.293& 43/10  &2.8$\times$2.2 (-22.4)  &0.12&0.010&18.04 $\pm$ 0.01\\
Q1416(taper)& - &-  & -& - &- &5.5$\times$5.5 (1.6)  &0.19&-&-\\

Q2121+0052 &  21:21:59.04&+00:52:24.1 &2.3732 $\pm$ 0.00002 &2.373 & 48/11  & 2.1$\times$1.4 (-61.1) &0.11 &0.011&18.07 $\pm$ 0.01\\
Q2121(taper)&-&-&-&-&-&5.8$\times$5.2 (-64.3)&0.21&-&-\\

Q2123$-$0050 &21:23:29.46 &  $-$00:50:52.9&2.2807 $\pm$ 0.00006 &2.280 & 47/10   &2.4$\times$2.0 (63.2) &0.13&0.014&16.34 $\pm$ 0.01\\
Q2123(taper) &-&-&-  &- &-&5.6$\times$5.3 (77.7) &0.23&-&-\\

\hline

\end{tabular}}
$^{\rm (a, b)}$ These two columns are from \citet{Cai2018, Cai2019}. $i$ mag means the $i$-band magnitude of each QSO.\\
$^{\rm (c, d)}$ represent the noise level of the corresponding datacubes, as well as the continuum images of the ALMA 12m data.

\end{table*}

\begin{table*}

\centering
\movetableright=-1.09in
  \caption{Results of the Gaussian fits to the CO(4-3) emission line spectrum of the QSOs, CO companions and extended regions.}
  \label{tab:gauss_fit}
\scalebox{0.78}{
\begin{tabular}{cccccccccc}
\hline
\hline
Target ID &R.A. &Decl.&Component& $v$ $^{\rm (e)}$     & $FWHM$  & $I_{\nu}$ $^{\rm (c)}$ & $error$-$level^{\rm (d)}$ &$L^{\prime}_{\rm CO(4-3)}$ $^{\rm (a)}$ & $M_\mathrm{H_2}$ $^{\rm (b)}$\\
 &(J2000) &(J2000)&& [\kms]  & [\kms]  & [Jy \kms] &&[10$^{10}$ K \kms\ pc$^{2}$] &10$^{10}$ M$_{\odot}$ \\
\hline
Q0050&-&-&QSO&8 $\pm$ 5&382 $\pm$ 11&2.16 $\pm$ 0.12&&3.25 $\pm$ 0.12& 11.7 $\pm$ 0.5\\
&-&-&QSO (taper)&3 $\pm$ 6&384 $\pm$ 14&2.45 $\pm$ 0.15& 1.5 &3.69 $\pm$ 0.09& 13.3 $\pm$ 0.4\\
&00:50:20.84&+00:51:33.4&CO emitter\_A&-14 $\pm$ 23&480 $\pm$ 57&0.36 $\pm$ 0.04& -&0.54 $\pm$ 0.07&4.3 $\pm$ 0.5\\
&00:50:20.14&+00:51:32.4&CO emitter\_B&323 $\pm$ 13&174 $\pm$ 31&0.13 $\pm$ 0.02&- &0.20 $\pm$ 0.03&1.6 $\pm$ 0.2\\
 \hline
Q0052&-&-&QSO&1 $\pm$ 10&221 $\pm$ 24&0.25 $\pm$ 0.02&-&0.40 $\pm$ 0.03&1.4 $\pm$ 0.1\\
 &00:52:31.94&+01:40:50.4&CO emitter\_A&76 $\pm$ 8&443 $\pm$ 18&0.91 $\pm$ 0.05&&1.46 $\pm$ 0.08&11.7 $\pm$ 0.7\\
 &-&-&CO emitter\_A (taper)&120 $\pm$ 36&509 $\pm$ 84&1.26 $\pm$ 0.19& 1.8 &2.02 $\pm$ 0.29&16.2 $\pm$ 2.3\\ 
 \hline
Q0101&-&-&QSO&7 $\pm$ 1&151 $\pm$ 2&1.26 $\pm$ 0.07&-&2.21 $\pm$ 0.14 &8.0 $\pm$ 0.4\\
 \hline 
Q0107&-&-&QSO\_blue$^{\rm (f)}$&-86 $\pm$ 8&287 $\pm$ 18&0.36 $\pm$ 0.04&-&0.56 $\pm$ 0.06&2.0 $\pm$ 0.2\\
&-&-&QSO\_red&207 $\pm$ 8&192 $\pm$ 18&0.77 $\pm$ 0.06&-&1.20 $\pm$ 0.09&4.3 $\pm$ 0.3\\
 &01:07:35.76&+03:14:34.9&CO emitter\_A&-121 $\pm$ 48&548 $\pm$ 112&0.84 $\pm$ 0.16&-&1.31 $\pm$ 0.27&10.5 $\pm$ 2.1\\
 &01:07:38.03&+03:14:46.7&CO emitter\_B&593 $\pm$ 11&143 $\pm$ 25&0.25 $\pm$ 0.04&-&0.39 $\pm$ 0.07&3.1 $\pm$ 0.6\\
  \hline 
Q1227&-&-&QSO& 0 $\pm$ 12& 476 $\pm$ 28&0.64 $\pm$ 0.04&-&0.98 $\pm$ 0.07&3.5 $\pm$ 0.3\\
 \hline
Q1228&-&-&QSO&16 $\pm$ 11&460 $\pm$ 26&0.82 $\pm$ 0.06&-&1.20 $\pm$ 0.24&4.4 $\pm$ 0.9\\
&-&-&QSO\_Narrow (taper) & -21 $\pm$ 17  & 306 $\pm$ 47 & 0.64 $\pm$ 0.13&-&0.94 $\pm$ 0.19&3.4 $\pm$ 0.7\\
&-&-&QSO\_Broad (taper) & -346 $\pm$ 130&1166 $\pm$ 226&0.79 $\pm$ 0.18&-&1.17 $\pm$ 0.26&4.2$\pm$ 0.9\\
&-&-&Extended emission&-217 $\pm$ 47&313 $\pm$ 110&0.14 $\pm$ 0.04&&0.20 $\pm$ 0.06&1.6 $\pm$ 0.4\\
&12:28:25.42  &+31:28:49.0&Extended emission (taper) &-413 $\pm$ 45&429 $\pm$ 104&0.43 $\pm$ 0.09 &3.0&0.62 $\pm$ 0.13&5.0 $\pm$ 1.0\\
&-&-&CO emitter\_A&315 $\pm$ 79&616 $\pm$ 186&0.41 $\pm$ 0.11&&0.60 $\pm$ 0.16 &4.8 $\pm$ 1.3\\
&12:28:26.16&31:28:51.5&CO emitter\_A (taper)&544 $\pm$ 48&522 $\pm$ 113&1.03 $\pm$ 0.20&2.7&1.51 $\pm$ 0.29&12.1 $\pm$ 2.3 \\
 \hline
Q1230&-&-&QSO&-4 $\pm$ 5&292 $\pm$ 12&1.17 $\pm$ 0.07&&1.89 $\pm$ 0.09& 6.8 $\pm$ 0.4\\
&-&-&QSO (taper)&-10 $\pm$ 8&343 $\pm$ 18&1.48 $\pm$ 0.10& 2.5 &2.39 $\pm$ 0.09&8.6 $\pm$ 0.4\\
&12:30:36.20&+33:19:53.1&CO emitter&81 $\pm$ 3&349 $\pm$ 7&1.73 $\pm$ 0.09&-&2.78 $\pm$ 0.15&22.2 $\pm$ 1.2\\
 \hline
Q1416&-&-&QSO&-7$\pm$ 22 &267 $\pm$ 51&0.14 $\pm$ 0.02&&0.22 $\pm$ 0.03&0.8 $\pm$ 0.1\\
&-&-&QSO (taper)&44$\pm$ 24 &268 $\pm$ 57&0.19 $\pm$ 0.03&1.4&0.29 $\pm$ 0.05&1.0 $\pm$ 0.1\\
&14:16:17.53&26:49:03.4&CO emitter\_A&-864 $\pm$ 12&156 $\pm$ 28&0.10 $\pm$ 0.02&-&0.16 $\pm$ 0.03&1.3 $\pm$ 0.2\\
&14:16:17.06&26:48:58.2&CO emitter\_B&-621 $\pm$ 37&683 $\pm$ 87&0.39 $\pm$ 0.04&&0.61 $\pm$ 0.06&4.9 $\pm$ 0.5\\
&-&-&CO emitter\_B (taper)&-480 $\pm$ 41&853 $\pm$ 98&0.61 $\pm$ 0.07&2.7&0.96 $\pm$ 0.11&7.7 $\pm$ 0.8\\
&14:16:17.25&26:49:03.9&CO emitter\_C&73 $\pm$ 18&133 $\pm$ 41&0.06 $\pm$ 0.02&-&0.09 $\pm$ 0.03&0.7 $\pm$ 0.2\\
 \hline
Q2121&-&-&QSO&5 $\pm$ 7&212 $\pm$ 16&0.28 $\pm$ 0.02&&0.47 $\pm$ 0.03&1.7 $\pm$ 0.1\\
&-&-&QSO(taper)&-8 $\pm$ 9&180 $\pm$ 21&0.37 $\pm$ 0.04&2.0&0.62 $\pm$ 0.07&2.2 $\pm$ 0.3\\
 \hline
Q2123&-&-&QSO&-7 $\pm$ 17&487 $\pm$ 40&0.50 $\pm$ 0.05&-&0.78 $\pm$ 0.08&2.8 $\pm$ 0.3\\
  &21:23:29.88&$-$00:50:51.8&CO emitter\_blue&-272 $\pm$ 26&549 $\pm$ 61&0.39 $\pm$ 0.04&&0.61 $\pm$ 0.06&4.9 $\pm$ 0.5\\
   &-&-&CO emitter\_blue (taper)&-263 $\pm$ 48&760 $\pm$ 112&0.63 $\pm$ 0.09&2.4&0.98 $\pm$ 0.14&7.8 $\pm$ 1.1\\
  &21:23:28.98&$-$00:50:53.4&CO emitter\_red&285 $\pm$ 22&520 $\pm$ 54&0.41 $\pm$ 0.04&-&0.64 $\pm$ 0.06&5.1 $\pm$ 0.5\\
 \hline
 
\end{tabular}}

$^{\rm a}$ $L^{\prime}_{\rm CO(4-3)}$ = $3.25 \times 10^{-7} I_{v}$ $\nu_{obs}^{-2}$ $D_{L}^{2}$ $(1+z)^{-3}$ from \citet{Solomon_Vanden_Bout2005}. \\
$^{\rm b}$ $M_\mathrm{H_2}$ = $\alpha_{\rm CO}$ $L'_{\rm CO(1-0)}$ from \citet{Solomon_Vanden_Bout2005} with $\alpha_{\rm CO}=3.6$ M$_{\odot}$ (K \kms\ pc$^{2}$)$^{-1}$\\
$^{\rm c}$ Errors of $I_{\nu}$ include the uncertainties of the noise-weighted Gaussian fit and the absolute flux calibration (5$\%$) for ALMA.\\
$^{\rm d}$ $error$-$level$ = ($I_{taper}-I_{untaper}$)/($I_{taper_{error}}^{2}+I_{untaper_{error}}^{2}$)$^{1/2}$ which reflects how significant the differences between the tapered and untapered flux are.\\
$^{\rm e}$ The velocities are with respect to the CO(4-3) redshifts in Table \ref{tab:cube}, in the reference frame of the corresponding QSO.\\
$^{\rm (f)}$ The double Gaussian fitting from Fig.~\ref{fig:spectrum5}
\end{table*}

\begin{table}
\caption{Results of all the continuum emitters in each QSO fields (except Q2121)} 
\centering

\movetableright=-1.1in
  \label{tab:continuum}

 \scalebox{0.88}{
\begin{tabular}{cccccccccc}
\hline
\hline

Target ID &R.A. &Decl.& $Flux$ $^{\rm a}$\\
-& -&-&mJy\,beam$^{-1}$ \\
Q0050  &- &- &0.26\,$\pm$\,0.02 \\
Q0052  &- &- &0.09\,$\pm$\,0.01 \\
 continuum emitter 1 &00:52:33.92 &01:40:42.6 &0.15\,$\pm$\,0.01 \\
 continuum emitter 2 &00:52:34.57 &01:41:00.2 &0.09\,$\pm$\,0.01  \\
Q0101  &- &- &0.22\,$\pm$\,0.01 \\
 continuum emitter 1 &01:01;16.78 &02:02:21.2 &0.09\,$\pm$\,0.01 \\
 continuum emitter 2 &01:01:18.12 &02:02:13.9 &0.06\,$\pm$\,0.01 \\
Q0107  &- &- &0.13\,$\pm$\,0.01 \\
Q1227  &- &- &0.13\,$\pm$\,0.01 \\
Q1228  &- &- & 7.87\,$\pm$\,0.40 \\
Q1230  &- &- &0.13\,$\pm$\,0.02 \\
 continuum emitter 1 &12:30:36.20&+33:19:53.1 &0.14\,$\pm$\,0.02 \\
Q1416  &- &- &0.38\,$\pm$\,0.02 \\
 continuum emitter 1 &14:16:17.06&26:48:58.2 &0.15\,$\pm$\,0.01 \\
Q2123  &- &- &0.10\,$\pm$\,0.01 \\
 continuum emitter 1 &21:23:28.98&$-$00:50:53.4 &0.08\,$\pm$\,0.01 \\

\hline

\end{tabular}}

$^{\rm a}$ Errors of $Flux$ include the absolute flux calibration (5$\%$) for ALMA.\\

\end{table}

\section{Results}
\label{sec:catalog}  
We detect CO(4-3) line emission in the host galaxies of all the ten $z\sim$2 QSOs. Six of them (Q0050, Q1228, Q1230, Q2121, Q1416, Q2123) show extended CO emission on spatial scales of 15$-$100 kpc, and seven of them (Q0050, Q0052, Q0107, Q1228, Q1230, Q1416, Q2123) contain at least one companion (CO emitter) in each QSO field. To constrain the possibility that the line and continuum-detected galaxies in the QSO fields are interlopers at lower redshift, we checked the coordinates of these galaxies in the SDSS and Dark Energy Spectroscopic Instrument (DESI) Legacy survey database and did not find any corresponding optical detection. The spatial distribution and the characteristics of the CO(4-3) line emission from these QSOs and their companions are shown in Figs. 1-11, and will be discussed individually in Sect \ref{sec:individual}.

To determine the centroid of these QSOs and companions, we obtained the total intensity (moment-0) maps integrated across the velocity ranges of the line emission from the untapered datacubes. Then we extracted the 1D spectrum against the peak of the CO(4-3) emission in the QSOs, companions and the extended regions from the untapered and tapered datacubes after primary beam correction.

The results of the Gaussian fitting of the spectra are summarized in Table\,\ref{tab:gauss_fit}. The {\sl curve fit} Python package was used to perform the fitting.\footnote{https://docs.scipy.org/doc/scipy/reference/generated/\\scipy.optimize.curve\_fit.html} We chose the area, center and sigma in the Gaussian function as free parameters, and regarded the 1 / rms of each channel as the weighting, with rms the root-mean-square noise level. To estimate molecular gas masses, for the QSOs, we assume r$_{\rm 4-3/1-0}$ = $L^{\prime}_{\rm CO(4-3)}$/$L^{\prime}_{\rm CO(1-0)}=1$, given that \citet{Riechers2011} found that QSOs typically contain highly excited molecular gas with constant brightness temperature up to the mid-$J$ levels of CO. For the extended emission and companion galaxies we assume r$_{\rm 4-3/1-0}$ = $L^{\prime}_{\rm CO(4-3)}$/$L^{\prime}_{\rm CO(1-0)}=0.45$, which is in agreement with excitation values found in the CGM and companion galaxies in the Spiderweb \citep{Emonts2018}, as well as the general population of sub-millimeter galaxies \citep[see review by][]{Carilli_Walter2013}. The molecular gas mass M$_\mathrm{H_2}$ is derived from the estimated $L'_{\rm CO(1-0)}$ \citep{Solomon_Vanden_Bout2005}, assuming $\alpha_{\rm CO}$\,=\,3.6 M$_{\odot}$ (K \kms\ pc$^{2}$)$^{-1}$ \citep{Bolatto2013, Daddi2010, Genzel2010}. Table \ref{tab:continuum} summarized the continuum results.

\subsection{Individual QSOs}
\label{sec:individual}
\begin{figure*}
\centering
\includegraphics[width=0.87\textwidth]{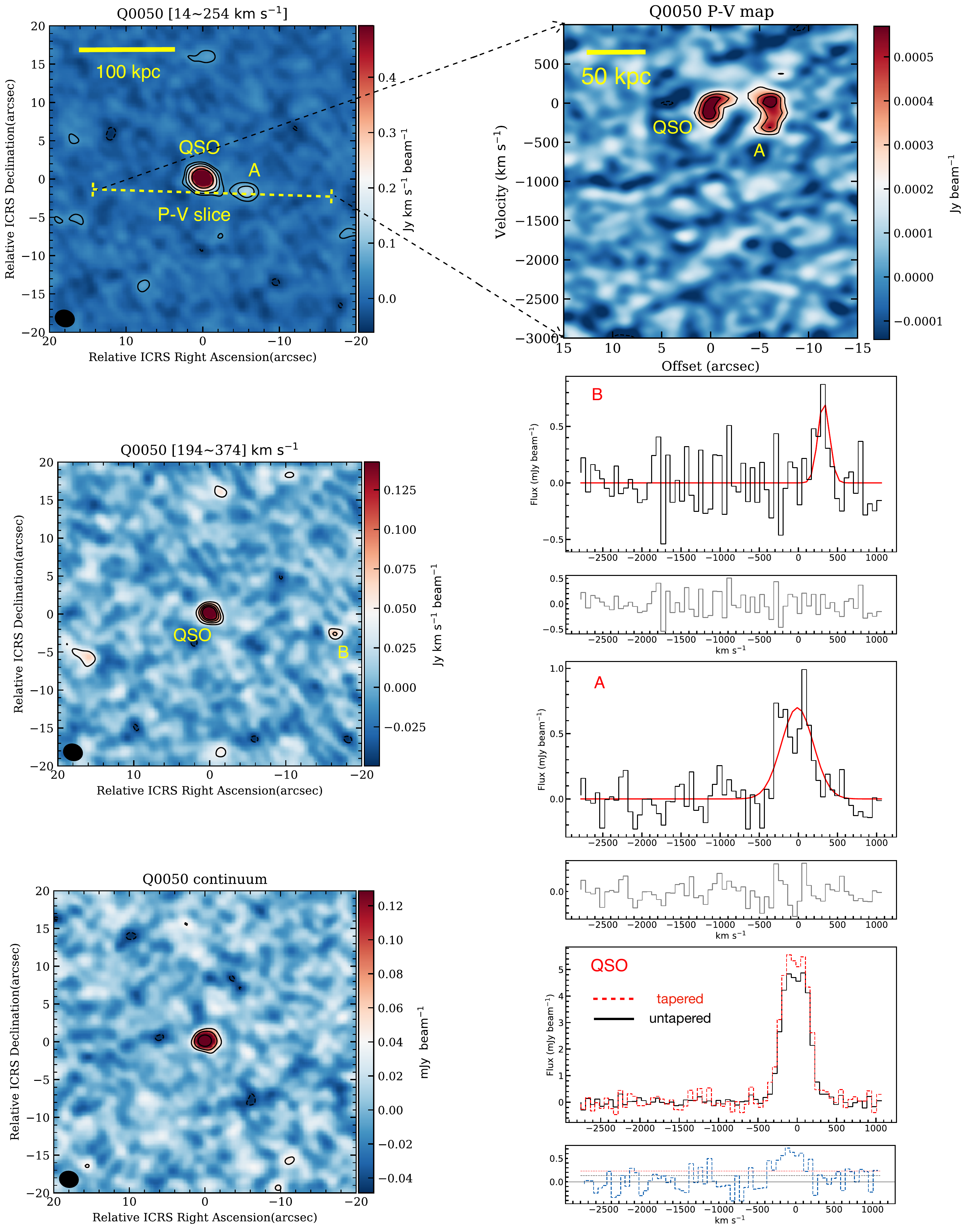}
\caption{The first column: Total intensity map collapsed from the 12m untapered CO(4-3) datacube of Q0050 integrated within $(14<v< 254)$ and $(194<v<374)$ \kms\ with 1$\sigma$ corresponding to 0.024 and 0.014Jy beam$^{-1}$ km s$^{-1}$, as well as the continuum image of Q0050 with 1$\sigma$ corresponding to 0.014 mJy beam$^{-1}$. The contour levels are (-3, 3, 6, 12, 24) $\sigma$, (-3, 3, 5, 7, 9) $\sigma$, (-3, 3, 6, 12) $\sigma$ from up to below. The second column: The Position-Velocity (P-V) map extracted along a 1-dimensional P-V slice, which location and direction is indicated in the left panel. For the PV-map, 1$\sigma$ equals 0.08 mJy\,beam$^{-1}$ channel$^{-1}$, and the contour levels are (-3, 3, 5, 7) $\sigma$. The three spectra below show the 1D spectra of the beam-integrated flux extracted from the 12m untapered datacube of Q0050 after primary beam correction, taken against the peak of the CO emission associated with the QSO, A emitter, and B emitter. The dashed red line shows the tapered result of the QSO and the dashed blue line in the bottom panel shows the offset between the tapered and untapered results of the QSO. The dashed flat red and black lines in the bottom panel show the noise level of the untapered and tapered datacube. The red lines show the Gaussian fitting and the grey lines in each bottom panel show the residuals.}
\label{fig:spectrum3}
\end{figure*}

{\bf Q0050+0051}: This QSO shows the brightest CO emission among our sample of 10 $z\sim2$ QSOs. The P-V map in Fig.~\ref{fig:spectrum3} reveals CO emission at v\,$\sim$\,+100 \kms\ that extends out to $\sim$20 kpc in the direction of the companion A, which is located at a projected distance of $\sim$45 kpc from the QSO. The P-V map further shows that, at the same velocity of +100 \kms, also some CO-emitting gas seems to stretch off the companion in the direction of the QSO. It is likely that the CO extensions are tidal debris from an interaction between the galaxies. The QSO field contains another companion B almost in the same direction as companion A at a projected distance of $\sim$135 kpc from the QSO. 

The continuum image taken with the ALMA 12m array reveals an unresolved point source associated with the QSO.

\begin{figure*}
\centering
\includegraphics[width=1.0\textwidth]{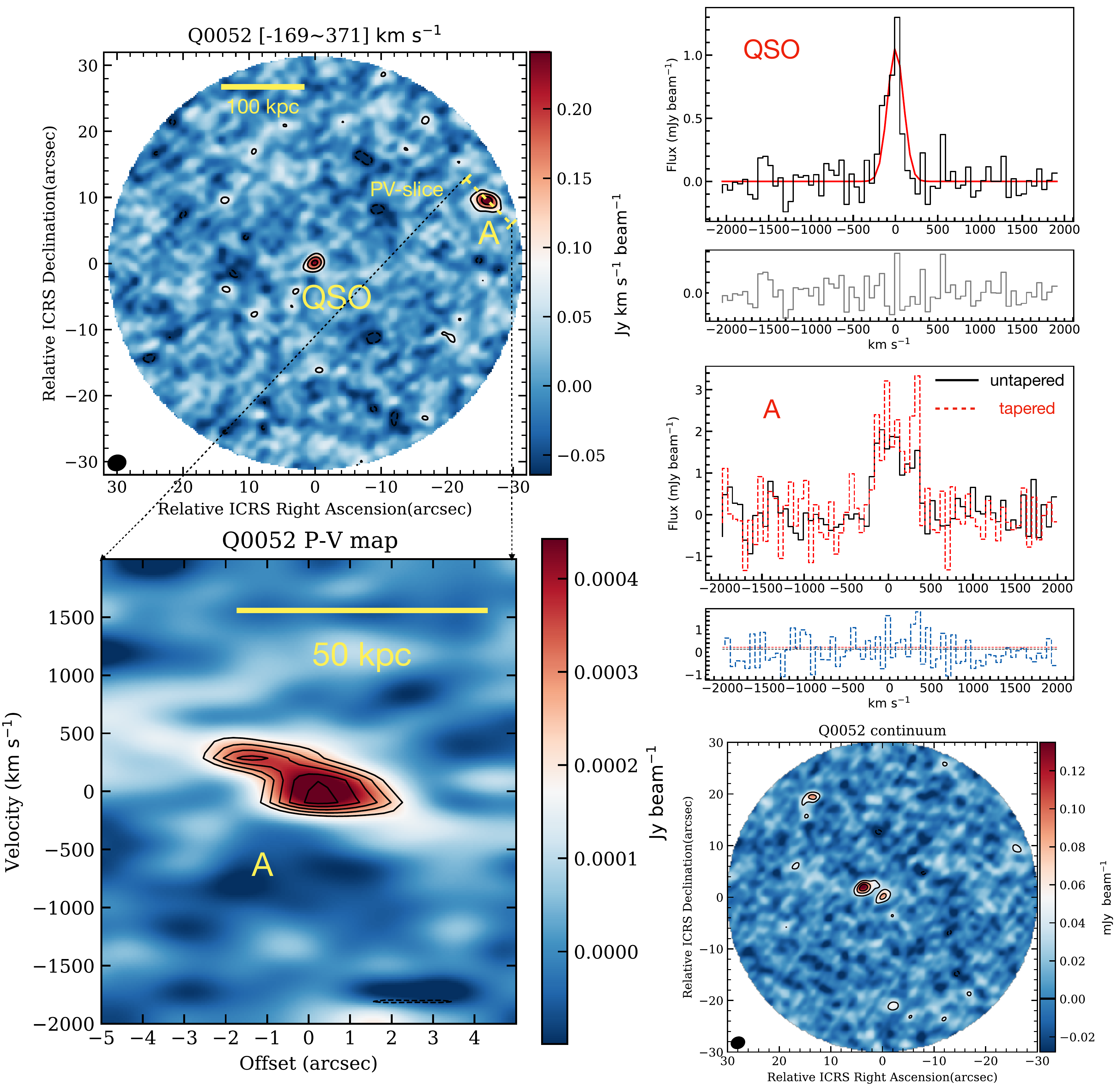}
\caption{The left side: The intensity maps collapsed from the 12m untapered CO(4-3) datacube of Q0052 integrated within $(-169<v< 371)$ \kms\ with 1$\sigma$ corresponding to 0.024 Jy beam$^{-1}$ km s$^{-1}$. The contour levels are (-3, 3, 6, 9) $\sigma$.  Also shown is the Position-Velocity (P-V) map extracted along the P-V slice-direction indicated in above panel with 1$\sigma$ equaling 0.075 mJy\,beam$^{-1}$ channel$^{-1}$ and contour levels corresponding to (-3, 3, 4, 5, 6, 7) $\sigma$. The right side: The black lines in the two panels show 1D spectra of the beam-integrated flux extracted from the 12m untapered datacube of Q0052 after primary beam correction, taken against the peak of the CO emission associated with the QSO and A emitter. The dashed red line shows the tapered result of A emitter and the dashed blue line in the bottom panel shows the offset between the tapered and untapered results of A emitter, The dashed flat red and black lines in the bottom panel show the noise level of the untapered and tapered datacube. The red lines show the Gaussian fitting and the grey lines in each bottom panel show the residuals. The panel below shows the continuum image of Q0052, with 1$\sigma$ corresponding to 0.013 mJy beam$^{-1}$. The contour levels are (-3, 3, 6, 9) $\sigma$.}
\label{fig:spectrum4}
\end{figure*}

{\bf Q0052+0104}: This QSO is unresolved, and its field contains a companion A (CO emitter) that is five times brighter in CO(4-3), located at a projected distance of $\sim$240 kpc from the QSO. This distant companion reveals extended CO emission across $\sim$40 kpc in NE-SW direction (Fig.~\ref{fig:spectrum4}).

The continuum image taken with the ALMA 12m array reveals an unresolved point source associated with the QSO and another two marginally resolved emitters at a projected distances of $\sim$16 and 190 kpc from the QSO.

\begin{figure*}
\centering
\includegraphics[width=1.0\textwidth]{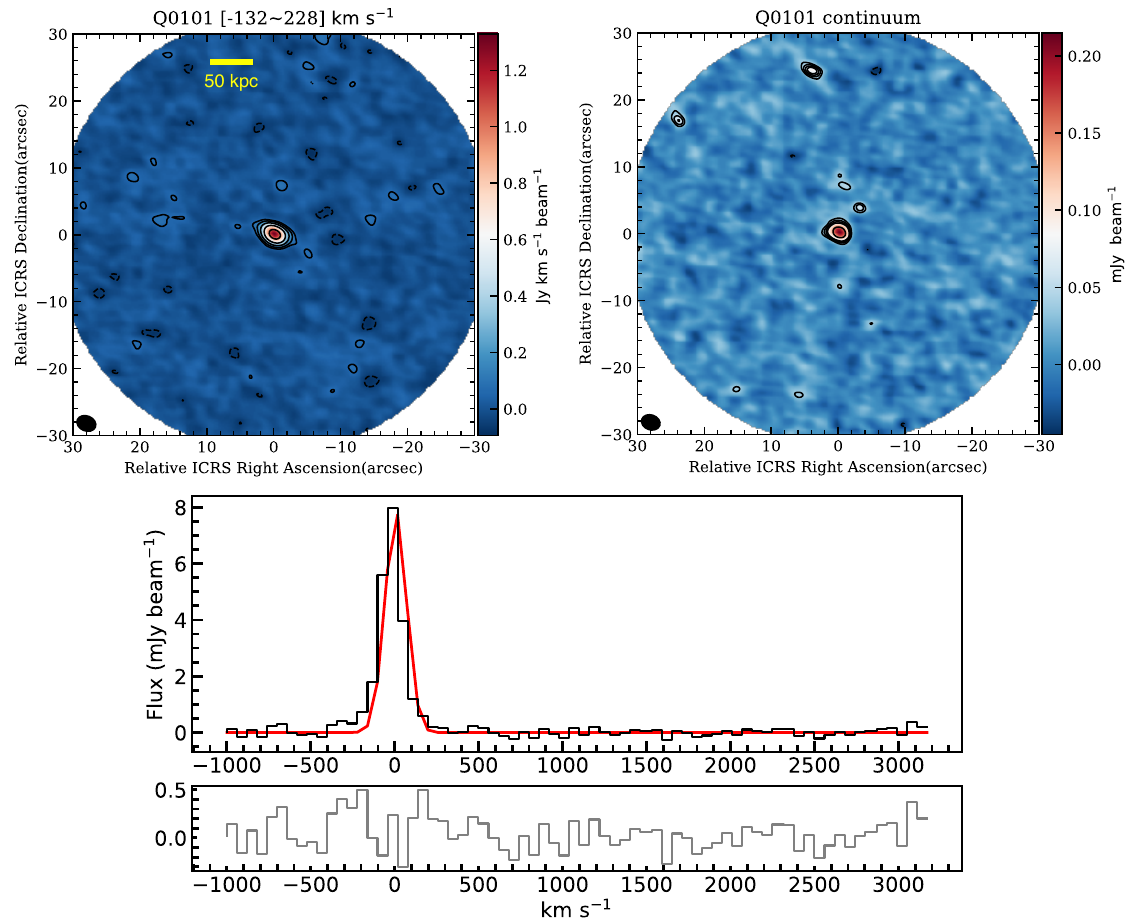}
\caption{{\sl Top}: Left: Total intensity map collapsed from the 12m untapered CO(4-3) datacube of Q0101 integrated within $(-132<v< 228)$ \kms\ and 1$\sigma$ is 0.021 Jy beam$^{-1}$ km s$^{-1}$. The contour levels are (-3, 3, 6, 12) $\sigma$. Right: The continuum image of Q0101 with 1$\sigma$ corresponding to 0.012 mJy beam$^{-1}$. The contour levels are (-3, 3, 4, 5, 6, 12) $\sigma$. {\sl Below}: 1D spectrum of the beam-integrated flux extracted from the 12m untapered datacube of Q0101, taken against the peak of the CO emission associated with the QSO. The red line shows the Gaussian fitting and the grey line in the bottom panel shows the residual.}
\label{fig:spectrum6}
\end{figure*}

{\bf Q0101+0201}: This QSO shows the second brightest CO emission in our sample and is unresolved in both CO(4-3) and continuum (Fig.~\ref{fig:spectrum6}). In addition, the CO(4-3) spectrum of the QSO shows an extremely small velocity dispersion (FWHM$\sim$151 \kms).  

The continuum image taken with the ALMA 12m array reveals three unresolved point sources, one is associated with the QSO and another two emitters are at projected distance of $\sim$190 and 230 kpc from the QSO.

\begin{figure*}
\centering
\includegraphics[width=1.0\textwidth]{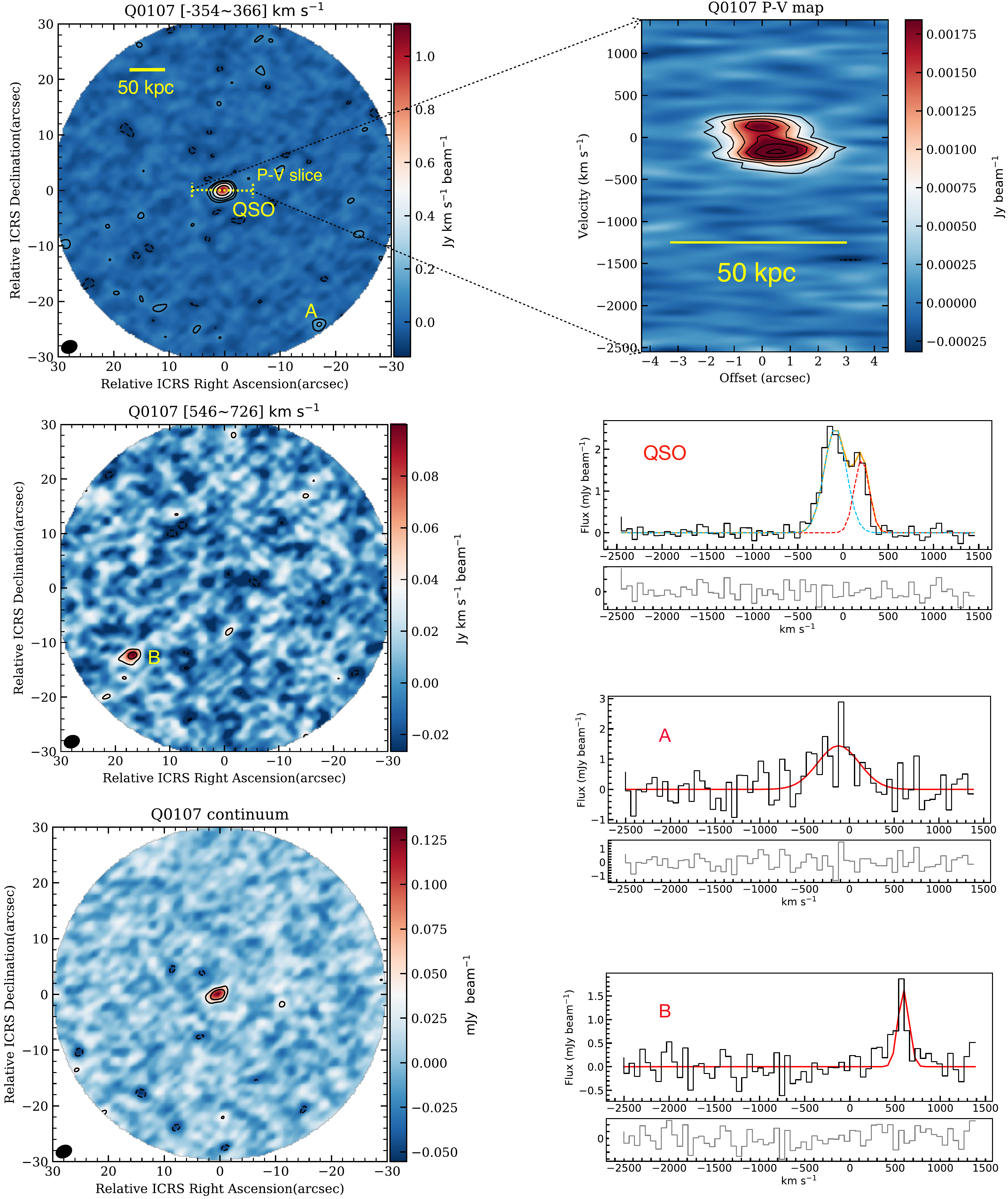}
\caption{{\sl Left}: Top and middle: Total intensity map collapsed from the 12m untapered CO(4-3) datacube of Q0107 integrated within $(-354<v< 366)$ and $(546<v<726)$ \kms\ with 1$\sigma$ corresponding to 0.028 and 0.014Jy beam$^{-1}$ km s$^{-1}$. Below: The continuum image of Q0107 with 1$\sigma$ corresponding to 0.013 mJy beam$^{-1}$. For all of the images, the contour levels are (-3, 3, 6, 12, 24, 48) $\sigma$. {\sl Right}: From top to bottom: 1D spectra of the beam-integrated flux extracted from the 12m untapered datacube of Q0101 after primary beam correction, taken against the peak of the CO emission associated with the QSO, A emitter and B emitter. The dashed blue and red lines show the double Gaussian fitting of the QSO, the red lines show the Gaussian fitting of A and B emitter, and the grey lines in each bottom panel show the residual.}
\label{fig:spectrum5}
\end{figure*}

\begin{figure}
\centering
\includegraphics[width=0.48\textwidth]{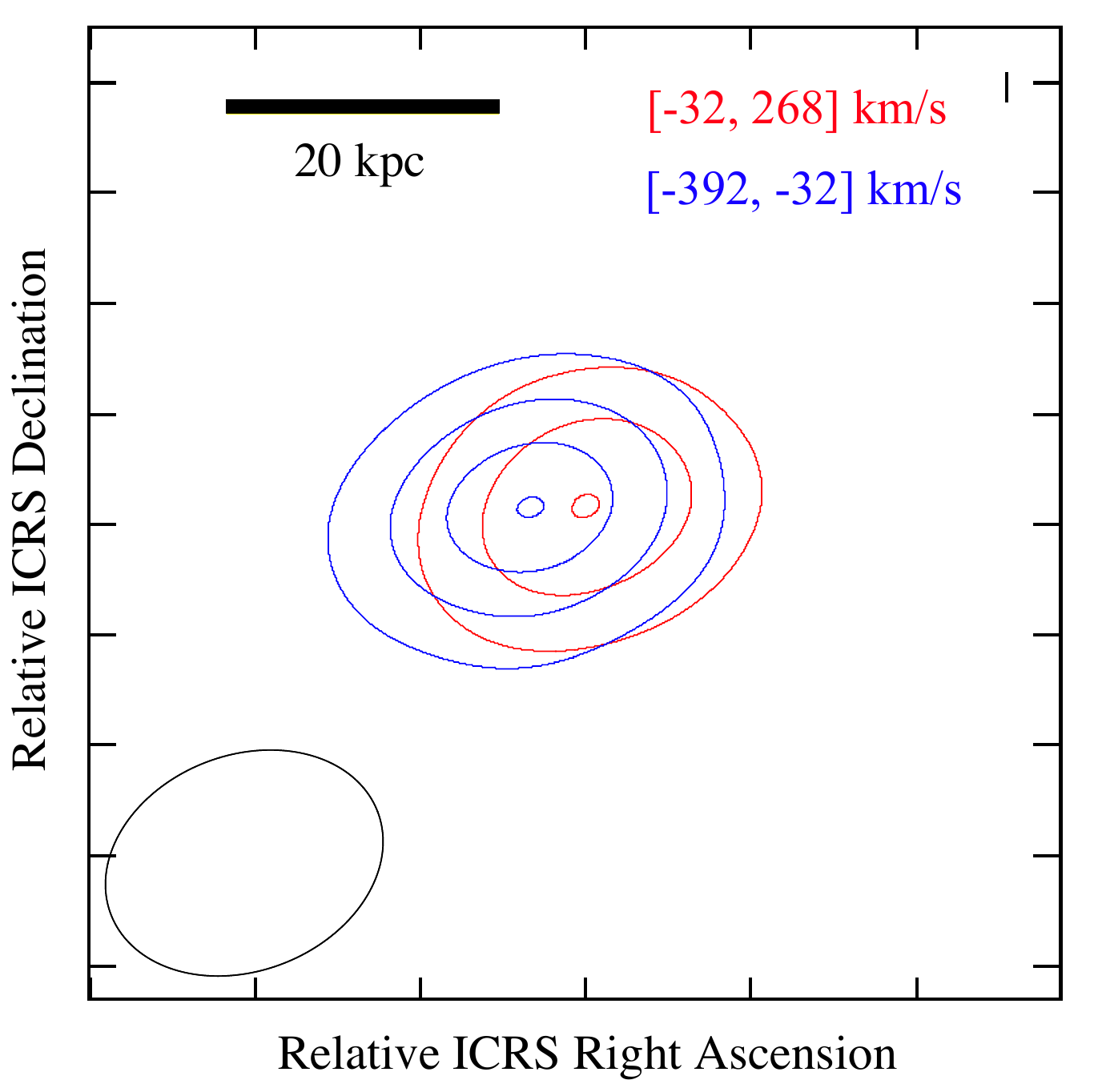}
\caption{Total intensity contours of the CO(4–3) emission against the QSO from Q0107 across the velocity ranges $(-392<v< -32)$ \kms(blue contours) and $(-32<v< 268)$ \kms(red contours). Contour levels are (9, 18, 27, 33) $\sigma$ for blue and (9, 18, 27) $\sigma$ for red, and 1$\sigma$ is 0.020 Jy beam$^{-1}$ km s$^{-1}$ for blue and 0.018 Jy beam$^{-1}$ km s$^{-1}$ for red.}
\label{fig:0107}
\end{figure}

{\bf Q0107+0314}: This source is kinematically resolved (Fig.~\ref{fig:0107}), and both the 1D spectrum and 2D P-V map of the QSO show a double peaked emission profile (Fig.~\ref{fig:spectrum5}). In addition, the field of this source contains two companions (CO emitters), one of them shows a small velocity dispersion (FWHM$\sim$143 \kms) at a projected distance of $\sim$160 kpc from the QSO, and the other is at a projected distance of $\sim$230 kpc from the QSO. 

The continuum image taken with the ALMA 12m array reveals an unresolved point source associated with the QSO.

\begin{figure*}
\centering
\includegraphics[width=1.0\textwidth]{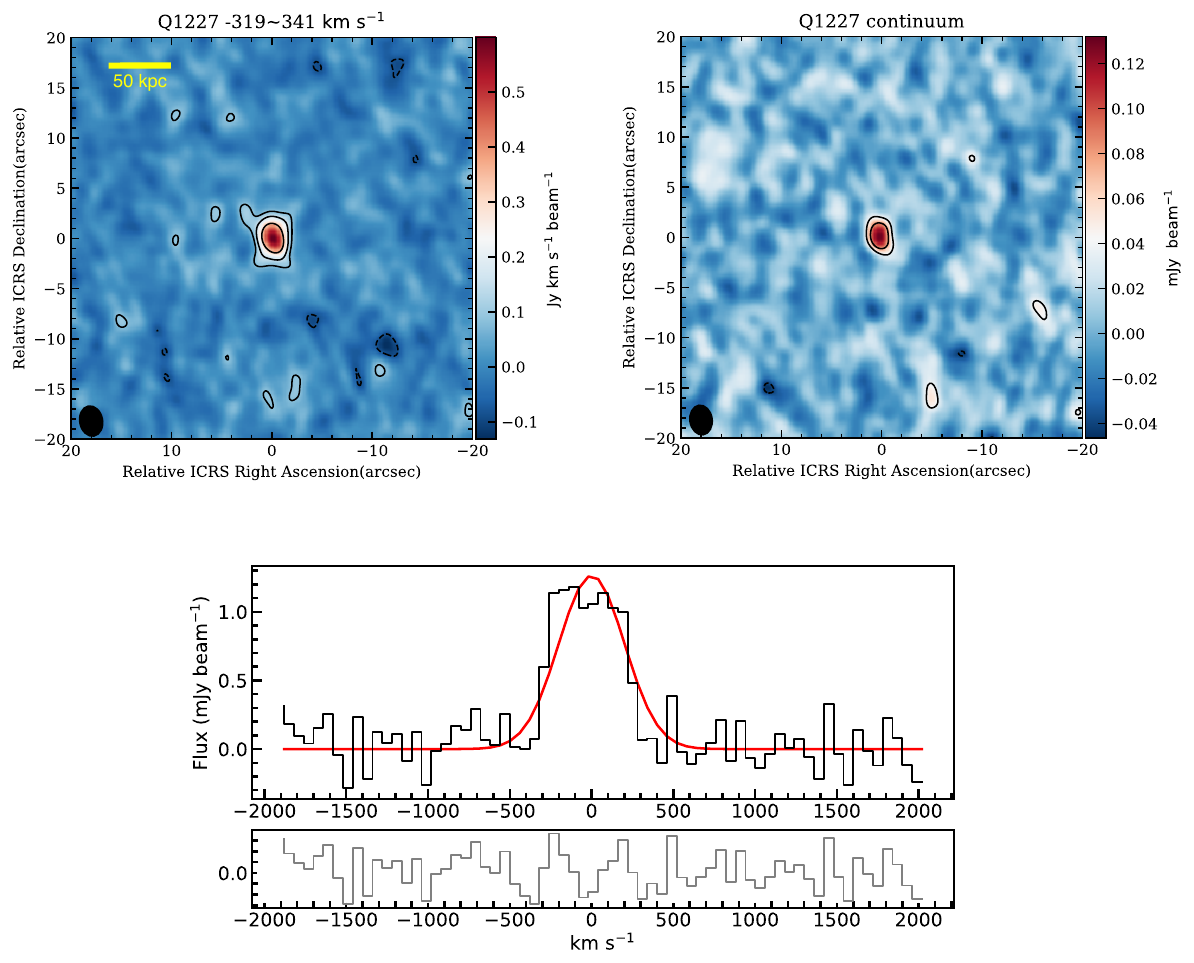}
\caption{{\sl Top}: Left: Total intensity map collapsed from the 12m untapered CO(4-3) datacube of Q1227 integrated within $(-319<v< 314)$ \kms and 1$\sigma$ is 0.027 Jy beam$^{-1}$ km s$^{-1}$. Right: The continuum image of Q1227 with 1$\sigma$ corresponding to 0.013 mJy beam$^{-1}$. For both of the images, contour levels are (-3, 3, 6, 12) $\sigma$. {\sl Below}: 1D spectrum of the beam-integrated flux extracted from the 12m untapered datacube of Q1227 against the peak of the CO emission associated with the QSO. The red line shows the Gaussian fitting and the grey line in the bottom panel shows the residual.}
\label{fig:spectrum7}
\end{figure*}

{\bf Q1227+2848}: This QSO is spatially unresolved in CO(4-3) (Fig.~\ref{fig:spectrum7}). The continuum image taken with the ALMA 12m array reveals an unresolved point source associated with the QSO.

{\bf Q1228+3128}: This object reveals the most extended CO reservoir and brightest continuum emission from the QSO in our ALMA observations. In addition, it is the only radio-detected quasar among our sample of 10 $z\sim2$ QSOs. 

\begin{figure*}
\centering
\includegraphics[width=1.0\textwidth]{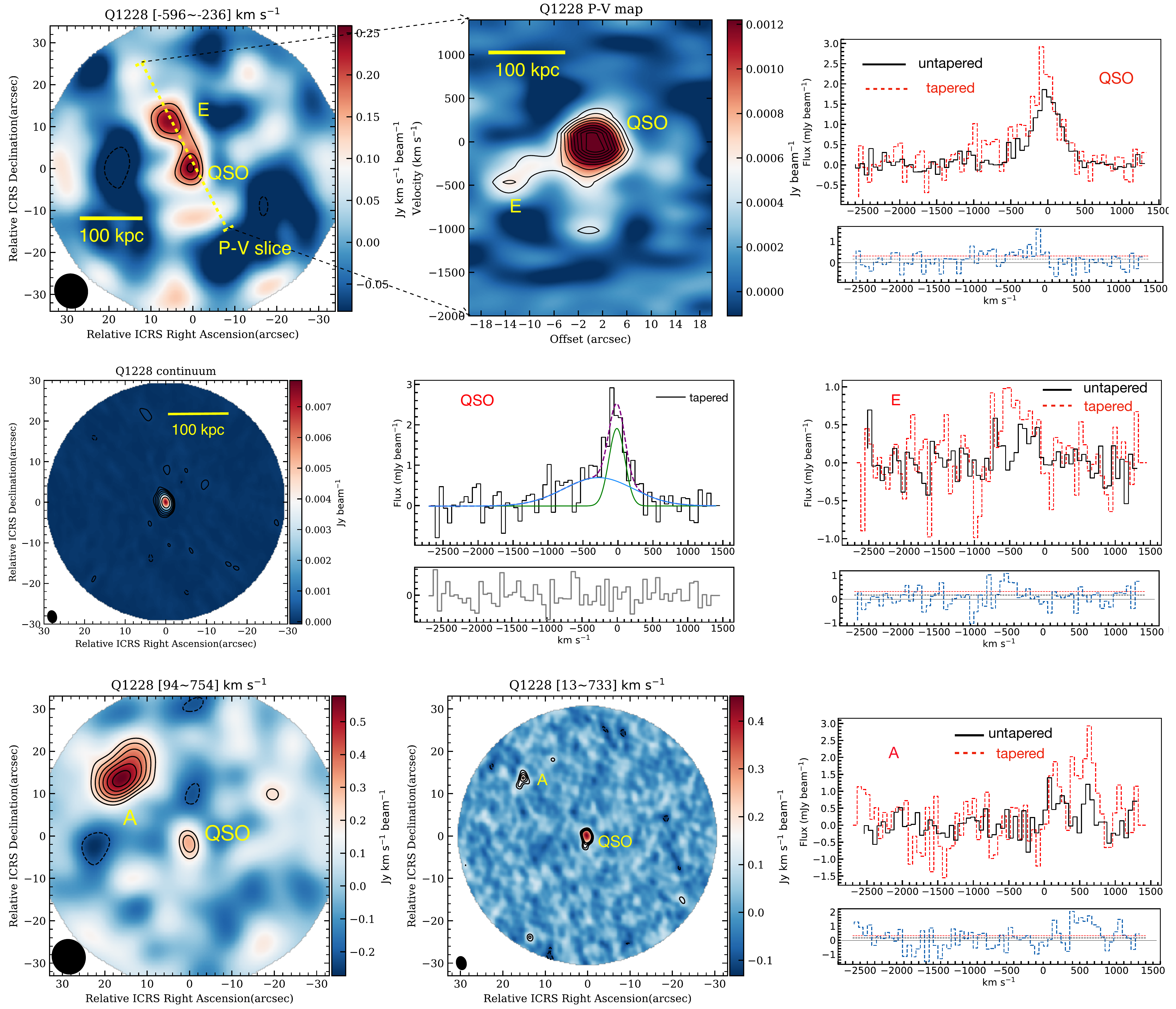}
\caption{{\sl First row}: Left: Total intensity map collapsed from the smoothed tapered CO(4-3) datacube of Q1228 integrated within $(-596<v< -236)$ \kms\ with 1$\sigma$ corresponding to 0.05 Jy beam$^{-1}$ km s$^{-1}$. Middle: Position-Velocity (P-V) map extracted along the P-V slice-direction indicated in the top-left panel with 1$\sigma$ equaling 0.18 mJy\,beam$^{-1}$ channel$^{-1}$. Right: The black and the dashed red line show the 1D spectra extracted from the 12m untapered and 7m+12m smoothed tapered datacube taken against the QSO, and the dashed blue line in the bottom panel show the offset between them. The dashed flat red and black lines in the bottom panel show the noise level of the untapered and tapered datacube. {\sl Second row}: Left: The continuum image of Q1228 with 1$\sigma$ corresponding to 0.043 mJy beam$^{-1}$. Middle: 1D spectrum of the beam-integrated flux extracted from the smoothed tapered datacube against the peak of the CO emission associated with the QSO. A double Gaussian fit has been applied to the line profile. The grey line in the bottom panel shows the residual. Right: The black and the dashed red line show the 1D spectra of the beam-integrated flux extracted from the 12m untapered and 7m+12m smoothed tapered datacube after primary beam correction, taken against the peak of the CO emission in the extended region, and the dashed blue line in the bottom panel show the offset between them. The dashed flat red and black lines in the bottom panel show the noise level of the untapered and tapered datacube. {\sl Third row}: Left: Total intensity map from the smoothed tapered datacube integrated within $(94<v<754)$ \kms with 1$\sigma$ corresponding to 0.067 Jy beam$^{-1}$ km s$^{-1}$. Middle: Total intensity map from the 12m untapered datacube integrated within $(13<v<733)$ \kms with 1$\sigma$ corresponding to 0.035 Jy beam$^{-1}$ km s$^{-1}$. Right: The black and the dashed red line show the 1D spectra extracted from the 12m untapered and 7m+12m smoothed tapered datacube after primary beam correction, taken against the companion A, and the dashed blue line in the bottom panel show the offset between them. The dashed flat red and black lines in the bottom panel show the noise level of the untapered and tapered datacube. In all panels, the contour levels are (-3, 3, 4, 5, 6, 7, 8, 9) $\sigma$.}
\label{fig:spectrum1}
\end{figure*}

Fig.~\ref{fig:spectrum1} reveals an extended reservoir of CO spreading more than 100 kpc in NE direction. The tapered spectrum of the QSO was fitted by a double Gaussian, a narrow component to fit the quiescent gas in the QSO, and a broad component which is missing in the untapered spectrum. The broad component has FWHM = 1166 km\,s$^{-1}$ which is significantly larger than what is typical for high-$z$ QSOs and Sub-Millimeter Galaxies \citep{Carilli_Walter2013} and could suggest the presence of a massive outflow \citep{Li2021}. The untapered and tapered spectra of the extended region were also fitted by single Gaussians. For the extended region, the flux of the untapered spectrum (0.14 $\pm$ 0.04 Jy \kms) only accounts for 33\% of the tapered spectrum (0.43 $\pm$ 0.09 Jy \kms). This means that 67$\%$ of the total flux consists of  low-surface-brightness emission from extended molecular gas, which was not recovered at sufficient S/N to be detected in our untapered 12m data. In that case, we derived the molecular gas mass of the extended region to be (5.0 $\pm$ 1.0) $\times$ 10$^{10}$ M$_{\odot}$.

The quasar field also contains a companion (A) located in NE direction at a projected distance of $\sim$160 kpc from the quasar. It shows a large offset between the flux of the tapered (1.03 $\pm$ 0.20 Jy \kms) and untapered spectra (0.41 $\pm$ 0.11 Jy \kms), which means almost 60\% of the flux is coming from outside the galaxy.

The continuum image taken with the ALMA 12m array reveals an unresolved point source associated with the quasar. 

\begin{figure*}
\centering
\includegraphics[width=1.0\textwidth]{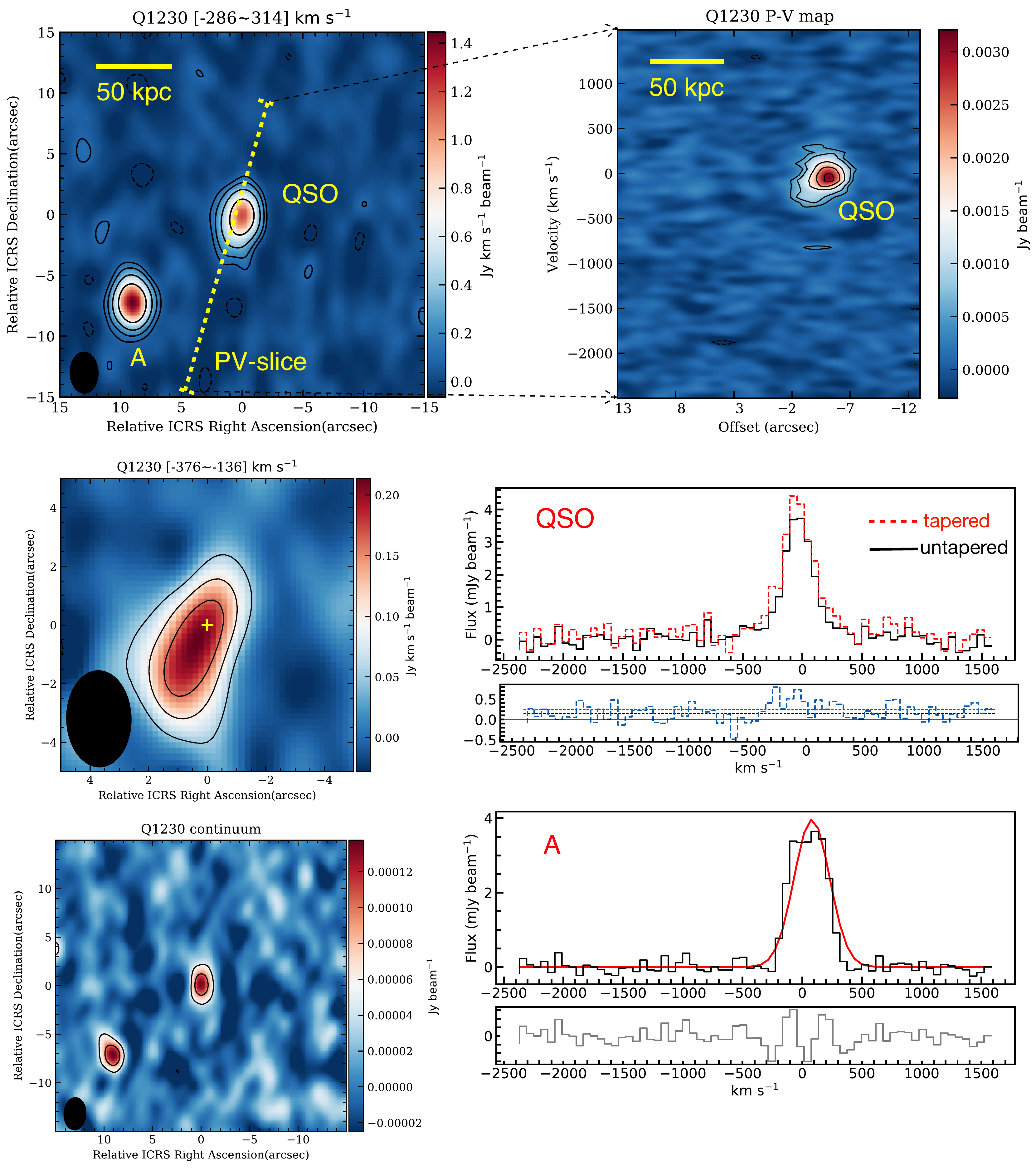}
\caption{The first column: The total intensity map collapsed from the 12m untapered CO(4-3) datacube of Q1230 integrated within $(-286<v< 314)$ \kms\ with 1$\sigma$ equaling 0.031 Jy beam$^{-1}$ km s$^{-1}$. The panel below shows the continuum image of Q1230, with 1$\sigma$ corresponding to 0.016 mJy beam$^{-1}$. For both panels, the contour levels are (-3, 3, 6, 12, 24) $\sigma$. The middle panel shows the zoom-in total intensity map within $(-376<v< -136)$ \kms, with 1$\sigma$ equaling 0.018 Jy beam$^{-1}$ km s$^{-1}$, and the contour levels are (3, 6, 9) $\sigma$ . The second column: The Position-Velocity (P-V) map extracted along the P-V slice-direction indicated in the left panel with 1$\sigma$ equaling 0.15 mJy\,beam$^{-1}$ channel$^{-1}$ and contour levels corresponding to (-3, 3, 5, 10, 15, 20) $\sigma$. And the panels below show the 1D spectra of the beam-integrated flux extracted from the 12m untapered datacube of Q1230 after primary beam correction, taken against the peak of the CO emission associated with the QSO and strong companion, respectively. The dashed red line shows the tapered result of QSO  and the dashed blue line in the bottom panel shows the offset between the tapered and untapered results of the QSO. The dashed flat red and black lines in the bottom panel show the noise level of the untapered and tapered datacube. The red line shows the Gaussian fitting and the grey line in the bottom panel shows the residual.}
\label{fig:spectrum8}
\end{figure*}

{\bf Q1230+3320}:  This QSO shows the third brightest CO emission in our sample and the P-V map in Fig.~\ref{fig:spectrum8} reveals extended CO emission associated with the QSO across $
\sim$15 kpc in SE direction. There is an even brighter unresolved CO emitter located at the SE direction at a projected distance of $\sim$80 kpc from the QSO (Fig.~\ref{fig:spectrum8}). 

The continuum image taken with the ALMA 12m array reveals two unresolved point sources associated with the QSO and companion.

\begin{figure*}
\centering
\includegraphics[width=1.0\textwidth]{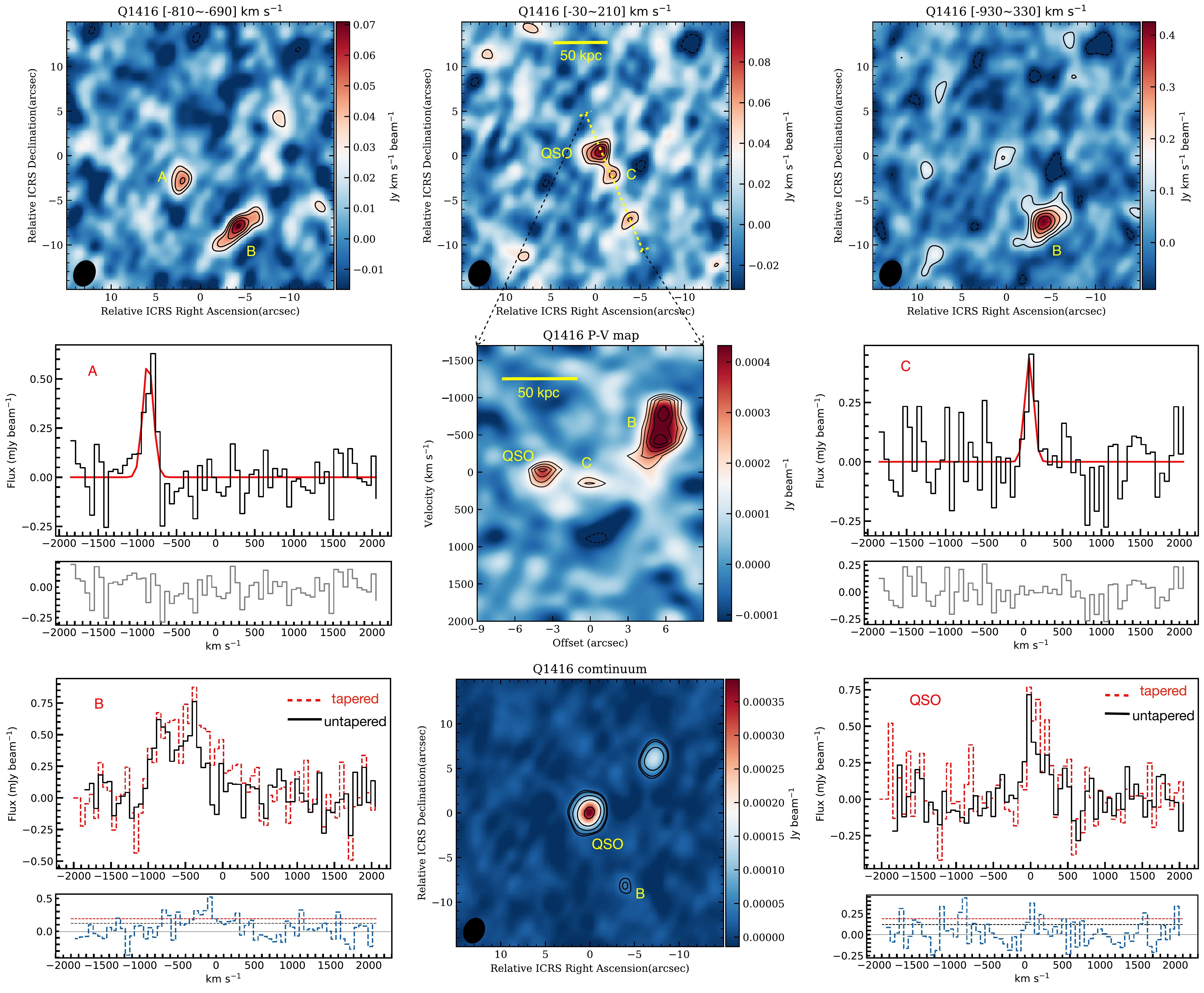}
\caption{The first row shows three total intensity maps collapsed from the 12m untapered CO(4-3) datacube of Q1416 integrated within $(-810<v<-690)$, $(-30<v<210)$ and $(-930<v<330)$ \kms, with 1$\sigma$ corresponding to 0.010, 0.014 and 0.033 Jy beam$^{-1}$ km s$^{-1}$. The contour levels are (-3, 3, 4, 5, 6) $\sigma$ for the first two maps and (-3, 3, 5, 7, 9, 11) $\sigma$ for the third map. The central panel shows the Position-Velocity (P-V) map extracted along the P-V slice-direction indicated in the above panel with 1$\sigma$ equaling 0.070 mJy\,beam$^{-1}$ channel$^{-1}$ and contour levels corresponding to (-3, 3, 4, 5, 6, 7) $\sigma$. The middle of the third row shows the continuum image of Q1416 with 1$\sigma$ equaling 0.010 mJy beam$^{-1}$ and contour levels corresponding to (-3, 3, 4, 8, 16, 32) $\sigma$. The other two panels in the second row shows the 1D spectra of the beam-integrated flux extracted from the 12m untapered datacube of Q1416 after primary beam correction, taken against the peak of the CO emission associated with the region of A and C, respectively. The red lines show the Gaussian fitting and the grey lines in each bottom panel show the residuals. For the other two panels in the third row, the black and dashed red line show the 1D spectra of the beam-integrated flux extracted from the 12m untapered and 12m+7m tapered datacube of Q1416 after primary beam correction, taken against the peak of the CO emission associated with the QSO and the B companion, respectively. And the dashed blue line in each bottom panel shows the offset between the tapered and untapered results, the dashed flat red and black lines in each bottom panel show the noise level of the untapered and tapered datacube.}
\label{fig:spectrum9}
\end{figure*}

{\bf Q1416+2649}: This QSO reveals widespread CO emission across $\sim$45 kpc in SW direction (Fig.~\ref{fig:spectrum9}). The QSO field contains at least two companions (CO emitters A and B), which are at a projected distance of $\sim$28 and 75 kpc from the QSO, respectively. Companion A shows a quite narrow FWHM$\sim$156 km\,s$^{-1}$, while companion B shows an extremely broad FWHM$\sim$853 km\,s$^{-1}$ and extended CO emission across $\sim$70/40 kpc in NW/NE direction. The Position-Velocity (P-V) map shows a plume of gas stretching off companion B in the direction of the QSO, and low-level emission in between companion B and the QSO. Component C is the the brightest part of the extended emission between B and the QSO with a narrow FWHM$\sim$133 km\,s$^{-1}$, but we note that it could also be a companion galaxy. We will further discuss this in Sect. \ref{sec:details}. 

For the QSO, the flux of the tapered spectrum (0.19 $\pm$ 0.03 Jy \kms) is roughly 36\% higher than that of the untapered spectrum (0.14 $\pm$ 0.02 Jy \kms), resulting in a difference equaling (0.05 $\pm$ 0.04 Jy \kms). Despite the large uncertainty, this is close to the flux of the extended reservoir C (0.06 $\pm$ 0.02 Jy \kms). For the broad-line companion B, the flux of the tapered spectrum (0.61 $\pm$ 0.07 Jy \kms) is roughly 56\% higher than that of the untapered spectrum (0.39 $\pm$ 0.04 Jy \kms).

The continuum image taken with the ALMA 12m array reveals two unresolved point sources associated with the QSO and another companion located in SW direction at a projected distance of $\sim$64 kpc from the QSO.

\begin{figure*}
\centering
\includegraphics[width=1.0\textwidth]{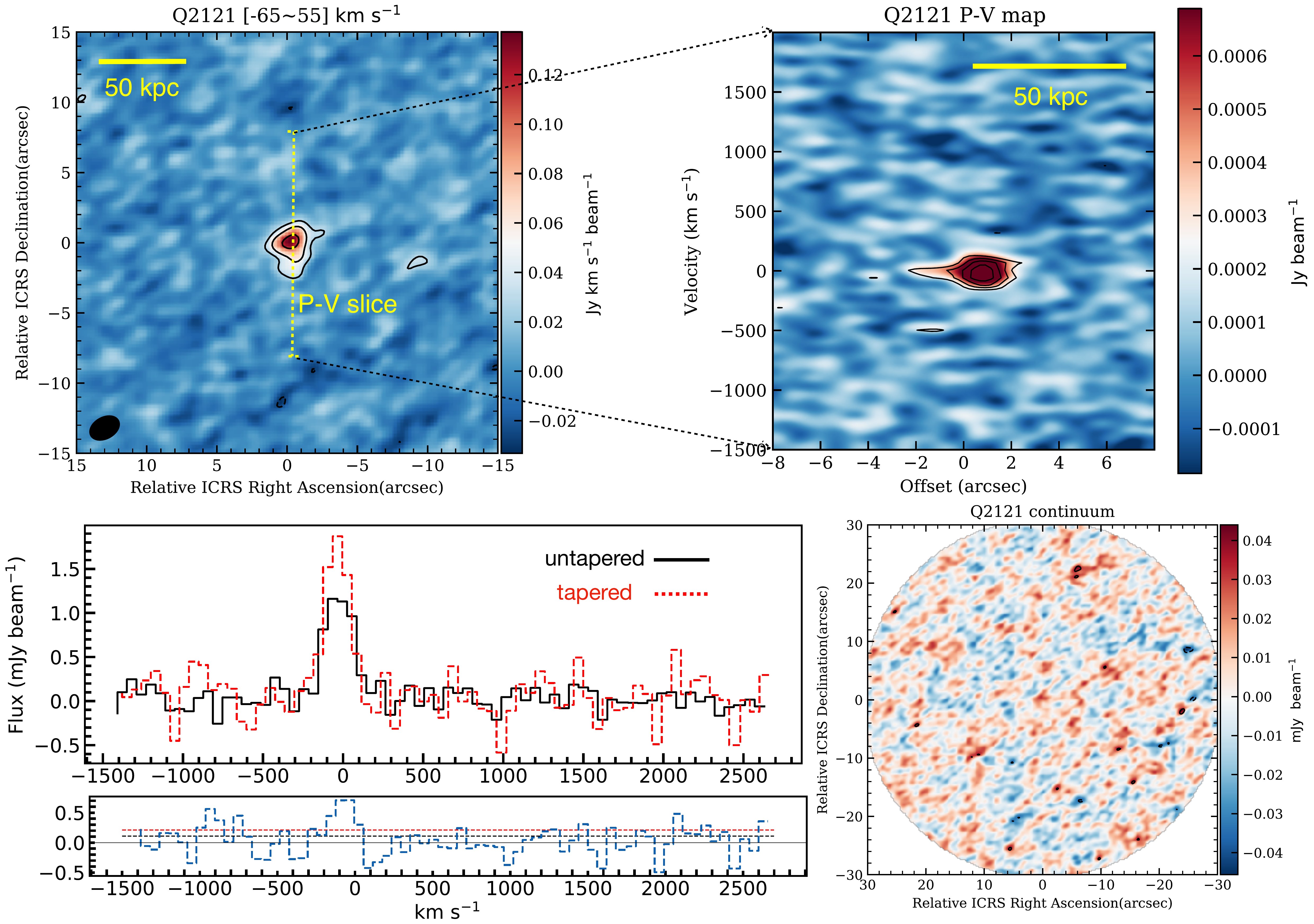}
\caption{{\sl Top}: Left: Total intensity map collapsed from the 12m untapered CO(4-3) datacube of Q2121 integrated within $(-65<v< 55)$ \kms and 1$\sigma$ is 0.01 Jy beam$^{-1}$ km s$^{-1}$. The contour levels are (-3, 3, 6, 12) $\sigma$. Right: Position-Velocity (P-V) map extracted along the P-V slice-direction indicated in the left panel with 1$\sigma$ equaling 0.11 mJy\,beam$^{-1}$ channel$^{-1}$. The contour levels are (-3, 3, 5, 7, 9) $\sigma$.{\sl Below}: Left: The black and dashed red line show the 1D spectra of the beam-integrated flux extracted from the 12m untapered and 12m+7m tapered datacube of Q2121 against the peak of the CO emission associated with the QSO, and the dashed blue line in the bottom panel shows the offset between them, the dashed flat red and black lines in the bottom panel show the noise level of the untapered and tapered datacube. Right: The continuum image of Q2121 with 1$\sigma$ corresponding to 0.011 mJy beam$^{-1}$.}
\label{fig:spectrum0}
\end{figure*}

{\bf Q2121+0052}: This is another object which also shows significant extended emission around the QSO. Fig.~\ref{fig:spectrum0} reveals extended CO emission across $\sim$35 kpc in Southern direction. The flux of the tapered spectrum (0.37 $\pm$ 0.04 Jy \kms) is roughly 32\% higher than that of the untapered spectrum (0.28 $\pm$ 0.02 Jy \kms). Furthermore, Fig.~\ref{fig:uv} in Appendix B shows an analysis of the CO emission of the central QSO in the ($u$,$v$)-domain through plotting the ($u$,$v$) distance of the baselines against the real part of the visibility amplitudes. From the Gaussian fitting (red line in Fig.~\ref{fig:uv}), we derived a zero-baseline flux density = $1.96 \pm 0.24$ mJy and a FWHM = ($4.0 \pm 0.2$)$^{\prime\prime}$ for the CO emission of the central QSO in the image plane, which corresponds to a spatial extent $\sim 33 \pm 2$ kpc. Both the spatial extent and total zero-baseline flux are consistent with our corresponding results in the image plane from Fig.~\ref{fig:spectrum0}. We obtain the best ($u$,$v$) fit if we also include a constant function (dashed black line in Fig.~\ref{fig:uv}) representing a CO point source with a flux density = 0.12 mJy. There is no signal detected in the continuum image taken with the ALMA 12m array.

\begin{figure*}
\centering
\includegraphics[width=0.80\textwidth]{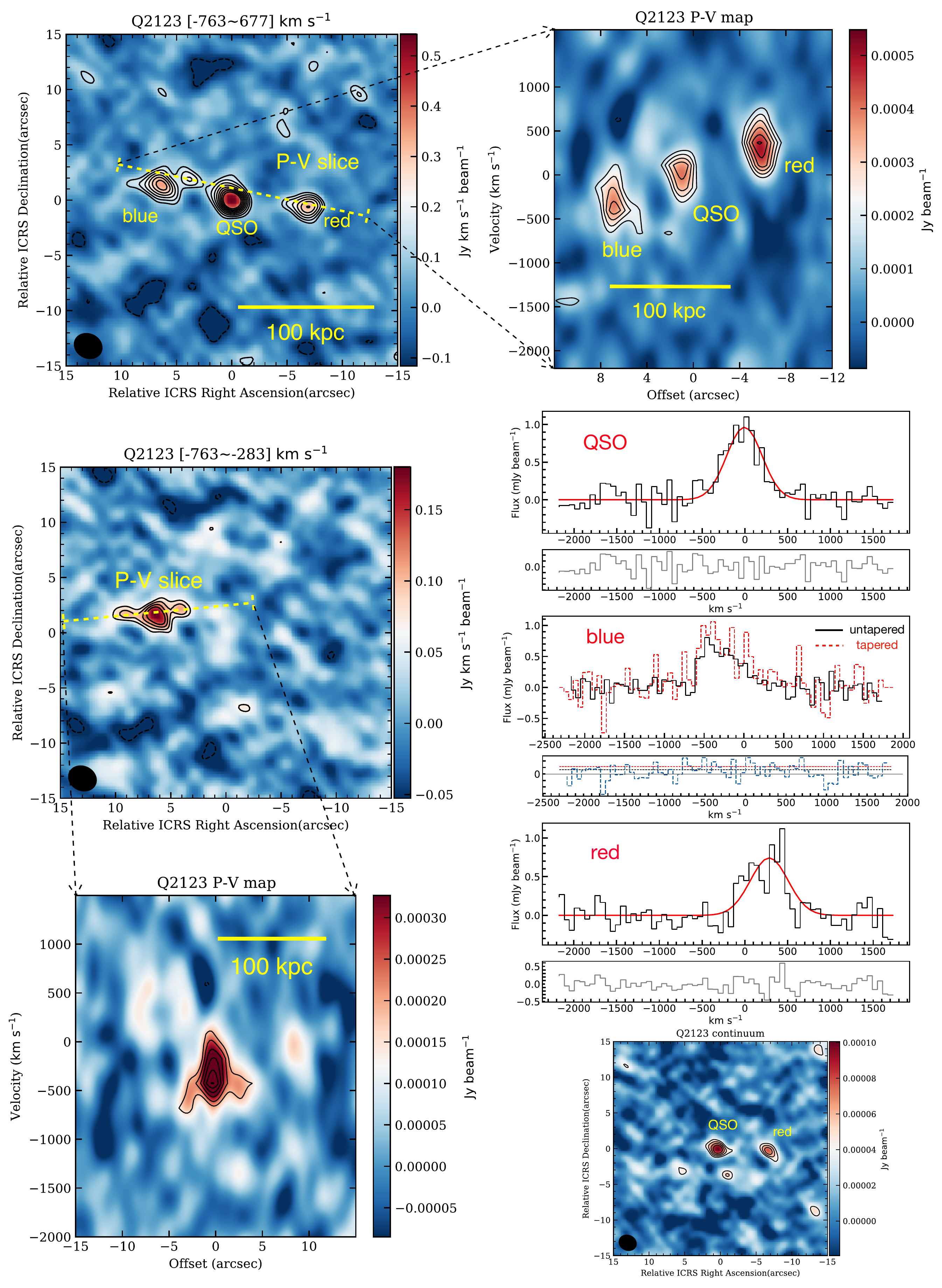}
\caption{{\sl Left}: Top and middle: Total intensity map collapsed from the 12m untapered CO(4-3) datacube of Q2123 integrated within $(-763<v< 677)$ and $(-763<v<-283)$ \kms, with 1$\sigma$ corresponding to 0.040 and 0.023Jy beam$^{-1}$ km s$^{-1}$. Below: Position-Velocity (P-V) map extracted along the P-V slice-direction indicated in the middle panel with 1$\sigma$ equaling 0.068 mJy\,beam$^{-1}$ channel$^{-1}$. {\sl Right}: Top: Position-Velocity (P-V) map extracted along the P-V slice-direction indicated in the left panel with 1$\sigma$ equaling 0.068 mJy\,beam$^{-1}$ channel$^{-1}$. Middle: 1D spectra of the beam-integrated flux extracted from the 12m untapered datacube of Q2123 after primary beam correction, taken against the peak of the CO emission associated with the QSO, blue and red side companions respectively. The dashed red line shows the tapered result of the blue side companion and the dashed blue line in the bottom panel shows the offset between the tapered and untapered results of the blue side companion, the dashed flat red and black lines in the bottom panel show the noise level of the untapered and tapered datacube. The red lines show the Gaussian fitting and the grey lines in each bottom panel show the residuals. Below: The continuum image of Q2123 with 1$\sigma$ corresponding to 0.014 mJy beam$^{-1}$. In all panels, contours start at 3$\sigma$ and increase with 1$\sigma$.}
\label{fig:spectrum2}
\end{figure*}

{\bf Q2123$-$0050}: This QSO is marginally resolved in CO(4-3), but there are also two bright companions (CO emitters) well aligned with the QSO both spatially and in velocity. Both of them are located at a projected distance of $\sim$56 kpc from the QSO. Fig.~\ref{fig:spectrum2} shows that the blue-side companion (i.e., the companion with lower CO velocities) reveals extended CO emission across $\sim$60 kpc in the EW direction. The flux of the tapered spectrum of the blue-side companion (0.63 $\pm$ 0.09 Jy \kms) is roughly 62\% higher than that of the untapered spectrum (0.39 $\pm$ 0.04 Jy \kms).

The continuum image taken with the ALMA 12m array reveals two unresolved point sources associated with the QSO and red-side companion. No continuum emission is detected for blue-side companion.

\subsection{Masses of the extended molecular gas reservoirs}

For many of our sources we find evidence for extended emission from the P-V plots. This is faint CO emission in the close vicinity of the stronger CO emitter (QSO or companion galaxy), but kinematically distinct. Nevertheless, due to beam dilution of the stronger emission, any flux estimates for these extended CO structures are highly uncertain, hence we do not estimate the mass from the P-V plots. 

The difference between the flux of the untapered and tapered results is another way of estimating the CO emission on large scales. For our sample sources, the differences between the flux of the untapered and tapered results suggest the mass of the extended molecular gas to be a few times 10$^{9}$ $\sim$ 10$^{10}$ M$_{\odot}$. However, Table\,\ref{tab:gauss_fit} (‘error-level’) shows that in most cases the flux-difference is small enough for mass estimates to be highly uncertain.  Therefore, we decided not to give estimates for the mass of the extended emission from the tapering analysis either. The exception is Q1228, where the extended reservoir is clearly well separated from the QSO and the uncertainty in the tapering analysis is lowest, and we derive (5.0 $\pm$ 1.0) $\times$ 10$^{10}$ M$_{\odot}$ for the extended region E.

\begin{figure}
\centering
\includegraphics[width=0.47\textwidth]{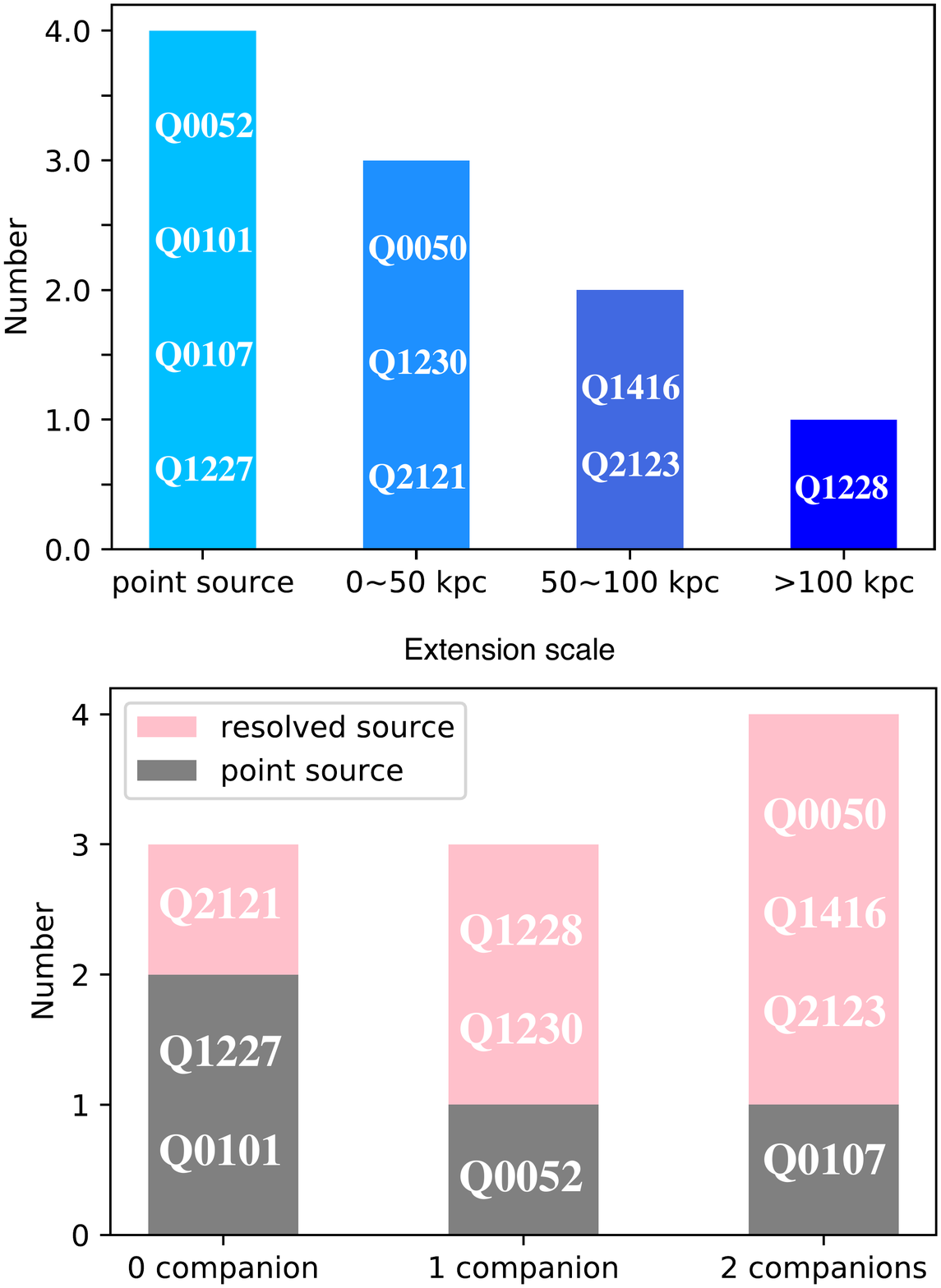}
\caption{These two panels show the statistical information from our sample about the scale of the extended CO emission and the number of the CO companions in each QSO field. The pink and grey columns represent the number of the CO companions detected in the fields of resolved and unresolved QSO.}
\label{fig:spectrum11}
\end{figure}

\begin{figure*}
\centering
\includegraphics[width=1.0\textwidth]{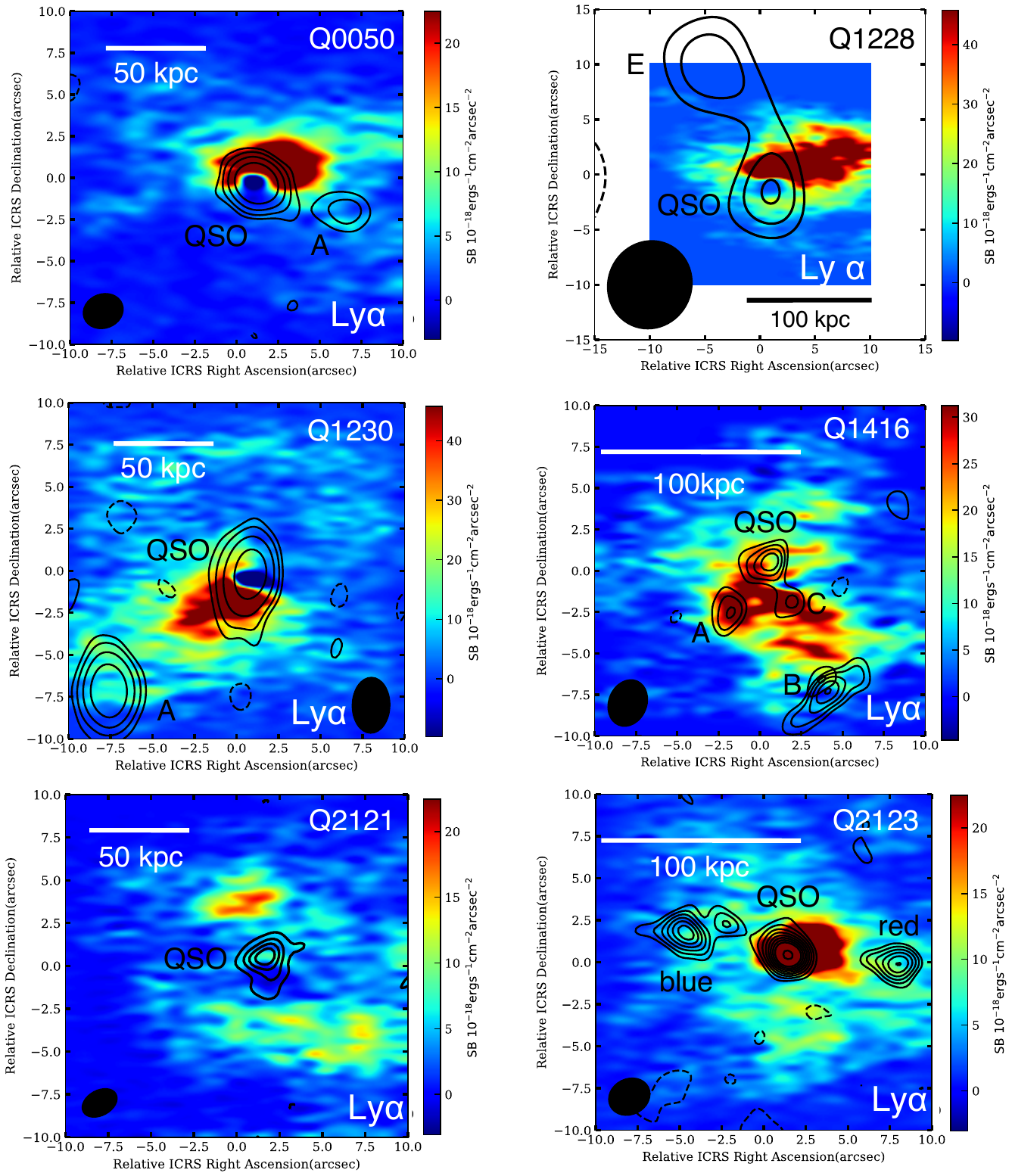}
\caption{For Q0050, Q1228, Q1230, Q1416, Q2121, Q2123, the black contours show the total intensity CO(4-3) map and the background shows the "Optimally extracted" Ly$\alpha$ nebula image from PSF- and continuum-subtracted KCWI data cubes, with 1$\sigma_{SB}$ equaling 5 $\times$ 10$^{-19}$ erg s$^{-1}$ cm$^{-2}$ arcsec$^{-2}$.}
\label{fig:spectrum13}
\end{figure*}

\begin{figure}
\centering
\includegraphics[width=0.475\textwidth]{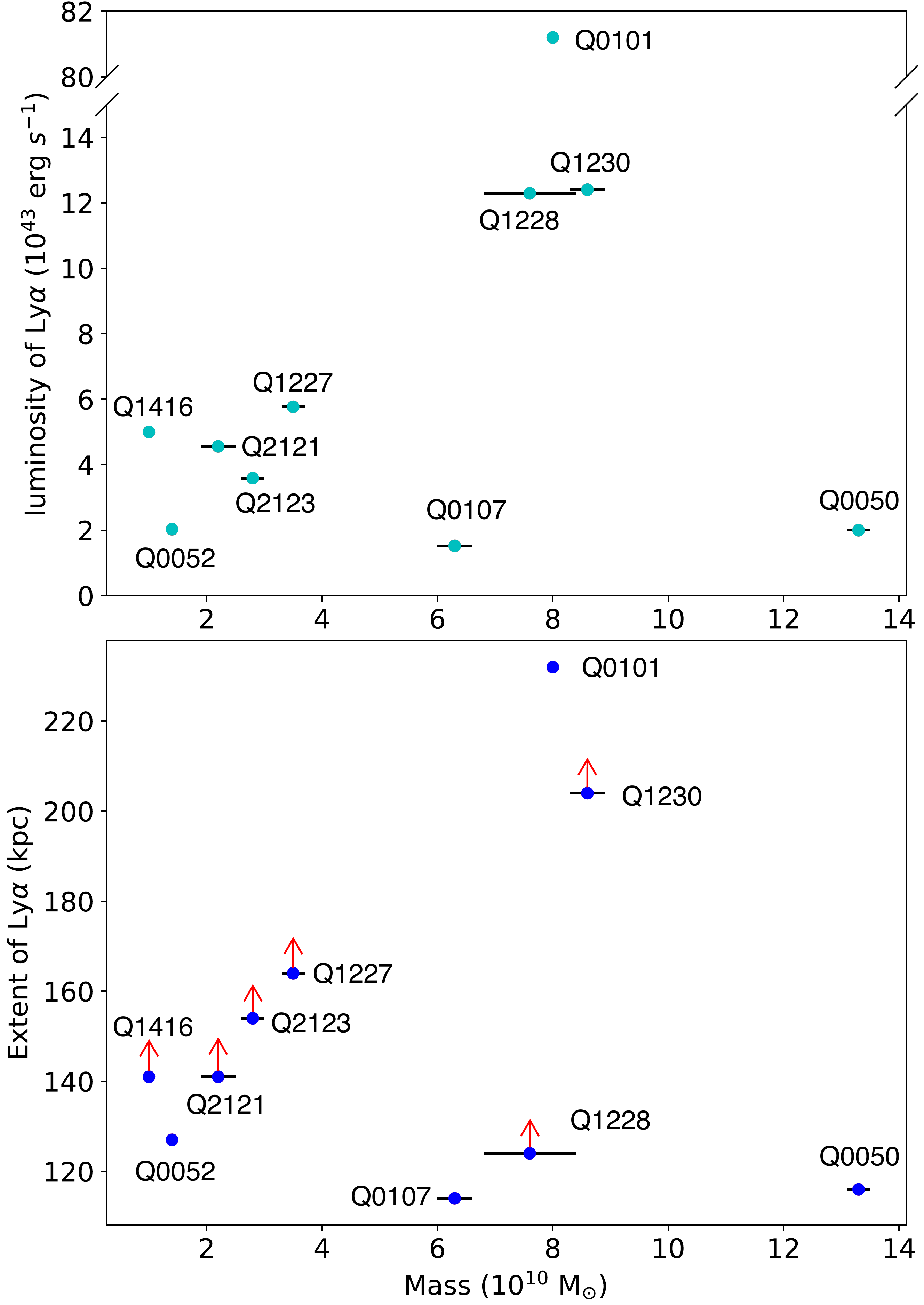}
\caption{Plots of molecular gas mass in the quasar host galaxy vs the luminosity and extent of Ly$\alpha$ nebulae \citep{Cai2019} for each target. The black bar shows the error, and the red arrow in the bottom panel shows the lower limit to the extent of the Ly$\alpha$ nebulae (see table.1 in \citealt{Cai2019}).}
\label{fig:spectrum14}
\end{figure}

\section{Discussion}

\subsection{Extended CO emission associated with the ELANe}

Fig.~\ref{fig:spectrum11} shows that in our sample of 10 targets, six of them (Q0050 $\sim$30 kpc, Q1228 $\sim$100 kpc, Q1230 $\sim$30 kpc, Q1416 $\sim$70 kpc, Q2121 $\sim$35 kpc, Q2123 $\sim$60 kpc) reveal extended CO emission associated with the QSO directly or a bright quasar companion (CO emitter) embedded in the ELANe. The other four sources do not show any significant signs of extended emission within the ELANe. Fig. \ref{fig:spectrum11} also indicates that the presence of progressively larger molecular reservoirs is increasingly rare, with only three ELANe (30$\%$) showing molecular gas across scales $>$50 kpc, and only one ELANe (10$\%$) with a molecular reservoir spread across $>$100 kpc.

\subsubsection{Details of sources with extended CO emission}
\label{sec:details}

The most extended molecular reservoir is found around Q1228, along the same direction as a central outflow of molecular gas \citep{Li2021}. Q1228 is the only radio-loud QSO in our sample, with the small radio source \citep{Helmboldt2007} also aligned with the outflow and extended molecular reservoir. For the massive molecular outflow and 100 kpc extended cold halo gas of Q1228, we propose the scenario that the extended CO-emitting gas is formed when the propagating radio jet drives the outflow, and also enriches or shocks and cools pre-existing dusty halo gas, while the Ly$\alpha$ emission could trace infalling gas behind the quasar (see \citealt{Li2021} for details). 

The two ELANe with molecular gas across scales of 50-100 kpc are Q1416 and Q2123. For Q1416, the broad-line companion (B) reveals both extended CO emission aligned with the QSO (marked by the plume that is visible in the PV plot; Sect. \ref{sec:individual}) and significant elongated CO morphology perpendicular to the QSO direction. In addition, it shows an extremely broad FWHM ($\sim$853 km\,s$^{-1}$) for the CO line, as well as an unresolved 4$\sigma$ continuum detection. In previous studies, only a few high-$z$ submillimeter galaxies and quasars have been discovered to display such broad CO lines \citep{Fogasy2020, Genzel2003, Neri2003, Greve2005, Coppin2008, Polletta2011, Frayer2018, Arrigoni_Battaia2022}. In those cases, 'rotation of a massive system' or 'high turbulence from a hidden AGN or a late stage merger event' were regarded as two main reasons of such broad lines \citep{Fogasy2020, Gnerucci2011, Williams2014}. For our case (the broad-line companion B), the spectrum taken with the ALMA 12m array tentatively reveals two components which are not completely separated in velocity, while the tapered spectrum shows additional flux in what appears to be a faint broad wing on the red side. This could indicate a molecular medium experiencing high turbulence. Possibly companion B is formed by two merging galaxies along the NW-SE direction, which are also experiencing a tidal interaction with the QSO along the NE-SW direction. In addition, the Q1416 field contains another nearby companion A, as well as a component C that is aligned both spatially and kinematically with the gas in the QSO and companion B. Both component A and C show a quite narrow FWHM ($\sim$133/156 km\,s$^{-1}$) CO line, but non-detection in the continuum image. There is just a minor offset ($\sim$80 km\,s$^{-1}$) between the narrow component C and the systematical redshift of QSO, but the companion A is separated by about $\sim$800 km\,s$^{-1}$ from our systematical redshift. Nevertheless, it is still within the FWZI (full width at zero intensity) of companion B. Moreover, both companion A and component C match the brightest region of Ly$\alpha$ emission (see Fig.~\ref{fig:spectrum13}). We propose the scenario that A and B are companion galaxies that are interacting with the QSO, and component C is either part of gaseous tidal debris in between B and the QSO, or a separate small companion that is caught in this debris. \citet{Fogasy2022} detected a merger system at $z\sim3.6$ containing a quasar SDSS J160705+533558 associated with a CO(1-0) companion, which are separated by 800 km\,s$^{-1}$ in velocity and $\sim$16.8 kpc on projected distance. \citet{Wang2011} detected a gas-rich, major merging system at $z\sim6.2$; where the CO (2–1) line emission from quasar J1429+5447 is resolved into two distinct peaks separated by $\sim$6.9 kpc in projected distance and with no clear velocity offset between the two components. These studies could be similar to what we observe for the Q1416 field. An alternative scenario is that the narrow FWHM of the CO(4-3) lines for components A and C indicates that the CO and Ly$\alpha$ emission in these regions represents dynamically relaxed gas reservoirs in the core(s) of the potential well of the ELAN. This could resemble the case of MAMMOTH-I, for which CO(1-0) observations revealed a dynamically cold CGM (FWHM$\sim$85 km\,s$^{-1}$) in the core of the proto-cluster, likely as a result of the cooling of an enriched multiphase medium \citep{Emonts2019}. 

Q2123 is a triple system, with extended molecular gas associated predominantly with one of the companions of the QSO. The CO emitters on both the red and blue side are likely a companion galaxies, but the extended CO emitter on the blue side shows a non-detection in the continuum image. Both the asymmetric spectrum and the P-V map of the blue companion does not show any kinematic evidence that the extended CO is part of a rotating disk. This may indicate that we detect the extended gas associated with the region around the blue companion and between the blue companion and the QSO. The morphology of the multiple cold gas reservoirs in Q2123 and associated extended CO emission resembles the CO(1-0) observations of MAMMOTH-I \citep{Emonts2019} which revealed extended CO emission stretching up to $\sim$ 30 kpc into the CGM. The blue companion could perhaps be a merging galaxy, which tidal interaction with the QSO gives birth to the extended emission. 

In comparison, Q0050, Q1230 and Q2121 all reveal extended emission on a much smaller scale (about 15$-$30 kpc). For Q1230 and Q2121, the extended CO emission is almost aligned with the direction of the extended Ly$\alpha$ reservoirs, although the Ly$\alpha$ emission is seen on a larger scale (see Fig.~\ref{fig:spectrum13}). And for Q0050 and Q1230, the extended cold gas is stretched in the direction of their companions, which could indicate that the extended emission of Q0050 and Q1230 is formed through tidal interaction between the QSO and its companion galaxy, much like the cases of Q1416 and Q2123.

To distinguish the above mentioned different scenarios, deep optical imaging of the QSOs and galaxies, and potential diffuse starlight between them (as was observed in the ELAN of the Spiderweb Galaxy by \citealt{Hatch2008}), is indispensable. Therefore, future deep optical imaging can give us further insight into the nature of the cold gas.

\begin{table}
\caption{The position angle information of extended CO emission and corresponding CO companion.} 
\centering

\movetableright=-1.1in
  \label{tab:PA}

 \scalebox{0.9}{
\begin{tabular}{cccccccccc}
\hline
\hline

Target ID & PA of extended CO emission & PA of CO companion\\
-& [deg]&[deg]\\
Q0050&52.74 &72.86\\
Q1228&22.82&47.79\\
Q1230 &162.74 &130.17\\
Q1416 &27.8 &22.96\\
Q2123 &-80.66 &80.22\\

\hline

\end{tabular}}

\end{table}

\subsubsection{Comparison with previous studies}

In Sect. \ref{sec:intro}, we gave an overview of -sometimes contradictory- results in the literature on the presence and absence of extended molecular gas in the CGM of various types of high-$z$ galaxies. In this section, we compare the CO(4-3) results of our sample of ELANe with previous result on cold molecular gas across three other enormous Ly$\alpha$ nebulae, namely the Spiderweb \citep{Miley2006}, Mammoth-I \citep{Cai2017b}, and the Slug (UM287; \citealt{Cantalupo2014}). The Spiderweb is a radio galaxy while the other two system surround radio-quiet AGN, though the Slug has a faint radio-loud quasar companion.

The evidence of cold molecular gas extending into the circumgalactic medium in low-J CO(1-0) and \ci\ was found in Spiderweb and MAMMOTH-1 \citep{Emonts2016, Emonts2018, Emonts2019}, while higher transition line of CO emission like CO(3-2) is localized to proto-cluster galaxies for MAMMOTH-1 \citep{Q2021} and shows no signs of extended structure. The non-detection of the extended emission of higher transition lines was also supported by the CO(4-3) and CO(3-2) observations of UM287 (Slug nebula) at $z\sim2$ \citep{Decarli2021, Chen2021}. However, at the same time, the CO(4-3) line is extended in the Spiderweb \citep{Emonts2018}, and \cii\ emission was also detected and partially ascribed to the CGM \citep{DeBreuck2022}.

According to these previous studies of ELANe, not the radio-quiet MAMMOTH-1 and UM287, but the radio-loud Spiderweb Galaxy revealed emission from the extended higher CO transition in CGM. For our sample, only the radio-loud target, Q1228, contains an extended CO(4-3) reservoir spread more than 100 kpc. In addition, compared to the huge ($\sim$70 kpc) cold gas halo of the Spiderweb, and also the $\sim$200 kpc CO(3-2) structure around the $z\sim$2.2 quasar cid\_346 \citep{Cicone2021}, the morphology of Q1228 is more directional, which resembles alignments previously seen between CO reservoirs and radio jets on scales of tens of kpc in high-z radio galaxies (HzRGs) \citep{Klamer2004, Emonts2014, Falkendal2021}. This may indicate a link between the inner radio-jet and the large-scale extended cold gas.

For the other five radio-quiet targets with extended CO(4-3) emission, four of them show extended CO(4-3) predominantly in the direction of their companion galaxies (see Table\,\ref{tab:PA}), and the scale of the extended cold gas also decreases to tens kpc. We argue that this extended CO-emitting gas is probably related to tidal debris from galaxy interactions. These extended CO reservoirs detected in the higher CO transition support enrichment of the CGM and potentially contribute to widespread star formation around massive galaxies and protoclusters. However, our results show no evidence for diffuse molecular gas spread across the extent of the Ly$\alpha$ nebulae from the CO(4-3) line.

Moreover, a substantial fraction of our sources (4/10) show no extended CO(4-3) across the nebulae. These first results from the SUPERCOLD-CGM survey therefore link the seemingly ambiguous results from the literature, where some galaxies -in particular high-$z$ radio galaxies- show evidence for extended molecular gas in the CGM,
%extended molecular gas in the CGM is frequently present in galaxy halos -in particular when considering high-$z$ radio galaxies-, 
while in other cases such extended molecular gas is notably absent.

\subsection{ Q0107: A massive rotating disk}

Observationally, disks have been identified via CO, H$\alpha$ and \cii\ spectroscopy at $z>2$ \cite[e.g.,][]{Genzel2003, Forster2009, Price2016, Genzel2017, Neeleman2020, Dannerbauer2017, Banerji2021, Fogasy2020}. Here we report a target in our sample, Q0107, which shows evidence for a massive CO disk associated with the QSO (see Appendix.A, Fig.~\ref{fig:optical}). Fig.~\ref{fig:0107} and Fig.~\ref{fig:spectrum5} show an offset $\sim$ 4 kpc and $\sim$ 293 km\,s$^{-1}$ between the double peak emission in spatial and frequency dimension, respectively. Although the CO signal is kinematically resolved, the spatial shift of the signal-peak is significantly smaller than the beam-size. Therefore, we have to rule out the positional errors, such as those introduced by phase instabilities, which could influence our results. The positional accuracy of the CO emission peaks in the intensity maps of Fig.~\ref{fig:0107} is given by: $\delta\theta_{rms} \sim 0.5 \langle \Theta_{beam} \rangle (S/N)^{-1}$ \citep{Papadopoulos2008}. For $\Theta_{beam}$= 2.3$^{\prime\prime}$, the FWHM of the synthesized beam, and S/N = 27, the signal-to-noise ratio, we derive a uncertainty $\delta\theta_{rms}$ to be 0.043$^{\prime\prime}$, which is ten times smaller than the spatial separation of $\sim 0.5^{\prime\prime}$.

Using a custom software code, QUBEFIT \citep{2020Neeleman} \footnote{https://github.com/mneeleman/qubefit}, we have fitted the data with a rotating disk model. The best-fit model gives a position angle (PA) of ${-80.6^\circ}^{+3.6^\circ}_{-4.1^\circ}$, an inclination angle ($i$) of ${42^\circ}^{+8^\circ}_{-8^\circ}$, an inclination-corrected rotational velocity (${v}_{rot}$) of ${339}^{+71}_{-46}$ km\,s$^{-1}$, and an exponential scale length of the modelled emission (${R}_{d}$) of ${1.53}^{+0.17}_{-0.14}$ kpc. To estimate the dynamical mass, we assume that the gas is rotationally supported. With this assumption, the dynamical mass (${M}_{\rm dyn}$) of a system (in ${M}_{\odot}$) within a radius R (in kpc) is given by ${M}_{\rm dyn}(R)$ = 2.32$\times10^{5}{{v}_{rot}}^{2}$R \citep{Neeleman2020}. The radius R is chosen to be $\sim$ 2.22 kpc, which is half of the distance between the peaks of the blue and red component showed in Fig.~\ref{fig:0107}. Within this region, we derive a dynamical mass of ${M}_{\rm dyn}$ = ${5.9}^{+2.9}_{-2.3}\times10^{10}$ ${M}_{\odot}$ which is consistent with the molecular gas mass $M_\mathrm{H_2}$ = $(6.3\pm0.3)\times10^{10}$ ${M}_{\odot}$. 

\subsection{Ly$\alpha$ nebulae properties vs. cold molecular reservoirs}

Fig.~\ref{fig:spectrum14} shows how the luminosity and total extent of the Ly$\alpha$ nebulae relates to the H$_2$ mass of molecular gas in the quasar host galaxies. Among our sample, Q0101 shows the brightest Ly$\alpha$ emission and also an quite narrow CO(4-3) spectral line (FWHM$\sim$151 \kms) which may indicate that there is a disk that is observed almost exactly face-on. What we found may confirm the new results reported by \citet{Costa2022}, who modeled the extended Ly$\alpha$ emission around high-$z$ quasars and found that the less the quasar host-galaxy is inclined with respect to our line-of-sight, the brighter and more extended the nebulae are. 

In addition,  Q0050, Q0052 and Q0107 are three of the dimmest and least extended Ly$\alpha$ nebulae \citep{Cai2019}, but contain masses of molecular reservoirs spread across the full range of our sample (see Fig.~\ref{fig:spectrum14}).  This is inconsistent with what \citet{Elgueta2022} reported. They explored many relations between the molecular gas and Ly$\alpha$ nebulae properties of a sample of nine $z\sim$3 quasars, and found a tentative trend that the more massive molecular reservoirs are associated with the dimmest nebulae. Moreover, in our sample, there are another three targets (Q0101, Q1228, Q1230), which contain relatively massive molecular reservoirs, but also show relatively large Ly$\alpha$ nebulae and bright Ly$\alpha$ emission.

\subsection{CO-emitting companions in the quasar fields}

The vast majority of QSO in our sample show companions in CO or continuum emission. In this Section, we will discuss the gas reservoirs associated with these companions, and investigate the over-densities in these QSO fields.

\subsubsection{Extended CO emission outside the ELANe}

For Q1228, there is a companion galaxy (A) located at a distance of $\sim$ 160 kpc from the quasar also revealing massive extended cold gas, almost in the same direction as the extended CO reservoir in region E, but at a higher velocity. The large amount of additional flux that is recovered after tapering the data might suggest that this could be a cold halo structure that resembles the huge CO reservoirs detected in the Spiderweb Galaxy \citep{Emonts2016} and cid$\_$346 \citep{Cicone2021}. Alternatively, a system similar to this companion is HAE229 ($z\sim$2.1), which is a star-forming galaxy located at a distance of $\sim$250 kpc from the central radio galaxy MRC 1138-262 in the Spiderweb protocluster, which also shows extended CO(1-0) emission across $\sim40$ kpc, likely as part of a large disk \citep{Dannerbauer2017}.

Q0052 reveals extended CO emission across $\sim$40 kpc associated with the bright companion that is at a distance of $\sim$240 kpc from the QSO, and $\sim$160 kpc outside the ELAN region. The P-V map in Fig.~\ref{fig:spectrum4} shows a significant kinematic gradient, which resembles both the extended CO(1-0)/(6-5) detection of a merger system associated with a $z\sim$3.4 submillimeter galaxy SMM J13120+4242 \citep{Frias2022} and the CO(1–0) detection of a large-scale disk in an ordinary, star-forming galaxy HAE229 in a protocluster at $z\sim$2.1 \citep{Dannerbauer2017}. Future deep optical imaging of the companion galaxy could help us to distinguish between these two possibilities.

\subsubsection{The QSO-Companion systems}

Fig.~\ref{fig:spectrum11} shows that in our sample, 70\% of the QSO targets contain at least one CO companion in their quasar field. These companions are generally located at distances of $\sim$25$-$240 kpc from the QSO. In detail, among our eleven CO companions, six show a distance from the QSO less than 100 kpc. 
Using the luminosity function of the CO(4-3) line with $L^{\prime}_{\rm CO(4-3)}$ ranging from 0.4 to 6.3 $\times$ 10$^{10}$ K \kms\ pc$^{2}$ at $z\sim$2.05$-$2.59 reported by \citet{Lenki2020}, we derive the predicted number of galaxies owning such bright CO(4-3) emission to be 0.03$\sim$0.001 in each of our fields of view (S $\sim$ 0.22 Mpc$^{2}$) with a depth of 400.8 Mpc ($\Delta z$ = 0.042) in sight-line direction. This means that at least 60\% of the QSOs (Q0050, Q0052, Q0107, Q1228, Q1230, Q2123) in our sample are located in significantly overdense regions. In addition, Q0052, Q0101, and Q1416 show at least one $>6\sigma$ continuum-detected source in the region where no CO was detected, at a distance of $\sim$16$-$190 kpc from the QSO. 

In previous studies, based on individual cases and small samples, interacting systems between the QSO and companion galaxies detected at submillimeter wavelengths at $z=2\sim6$ have been reported \cite[e.g.,][]{Ivison2008, Clements2009, Salome2012, Riechers2013, Fogasy2017, Fogasy2020, Diaz-Santos2018, Decarli2017, Trakhtenbrot2017, DOdorico2018, Neeleman2019, Stacey2022}. Besides, almost the same detection rate ($\sim$80\%) of sub-millimetre companions around the most luminous type I QSOs at $z\sim$2.4$-$4.7 was reported by \citet{Bischetti2021}; specifically, seven out of nine QSOs were found to have a bright CO or \cii\ emitter at about the same redshift and projected distances of $\sim$6$-$130 kpc from the QSOs. Using ALMA band 3 observations, \citet{Garcia2022} reported a blind search in the environments of 17 $z\sim$4 QSOs in CO(4-3), and they found that the predicted number of CO(4–3) emitting galaxies is 0.28, however five sources were detected in their survey. All of our results show that these luminous QSOs are preferentially located in high-density regions and the companions can be as bright as the QSO host galaxies. This provides important observational evidence that high-redshift QSOs reside in overdense regions. Moreover, recent FIR observations, taken with ALMA of Cosmic Dawn ($z\sim6$) QSO-companion systems, have revealed highly star-forming, infrared luminous host galaxies in the process of merging with nearby companions \citep{Decarli2017, Neeleman2019, Neeleman2021, Venemans2020}. Follow-up FIR observations of our Cosmic Noon ($z\sim2$) sample may help us to study the evolution of the properties of luminous QSOs host galaxies during cosmic time.

\section{Summary}

We report on ALMA/ACA Band\,4 observations of CO(4-3) line and continuum emission in a sample of 10 ultraluminous Type-I QSOs from the (SDSS)-IV/eBOSS database at $z\sim2.1-2.3$. These results enable us to perform the first statistical study on the role of the cold CGM in the evolution of active massive galaxies at Cosmic noon. Our main findings are summarised as follows:

(1). We detect CO(4-3) line emission in all QSOs among our sample (100\% detection rate), and the continuum emission of QSOs is detected in $\sim$90\% of our targets. Statistically, in 60\% of our targets, the CO(4-3) is extended on scales of 15 to 100 kpc in these ELANe. 

(2). The ELAN with the most extended CO(4-3) reservoir of $\sim$100 kpc (Q1228+3128) is the only radio-loud target in our sample, and the CO reservoir is found along the radio axis, which could indicate a link between the inner radio jet and cold halo gas spreading on a much larger scale. The other five radio-quiet ELANe show extended CO(4-3) predominantly in the direction of their companion galaxies. These extended cold reservoirs detected in higher CO transition support enrichment of the CGM and potentially contribute to widespread star formation around massive galaxies and protoclusters. However, there is no evidence for diffuse molecular gas spread across the extent of the Ly$\alpha$ nebulae from the CO(4-3) line. 

(3). We find a target in our sample (Q0107) showing significant evidence of a massive CO disk associated with the QSO. We derive a dynamical mass of ${M}_{dyn}$ = ${5.9}^{+2.9}_{-2.3}\times10^{10}$ ${M}_{\odot}$ which is consistent with the molecular gas mass $M_\mathrm{H_2}$ = $(6.3\pm0.3)\times10^{10}$ ${M}_{\odot}$. 

(4). Our ultraluminous QSOs are probably located in high-density environments, given the presence of at least one CO companion around 70\% of the QSOs. These CO companions show bright CO(4-3) emission, in some cases comparable to the QSOs, and are located at projected distances of $\sim25-240$ kpc from the QSO. In addition, Q0052, Q0101, and Q1416 show at least one $>$ 6$\sigma$ continuum source in the region where no CO is detected, at projected distance of $\sim16-190$ kpc from the QSO.

Our work gives us an insight into the crucial roles of both the cold CGM and companion galaxies in driving the evolution of central massive galaxies in the Early Universe. For the next step, the CO(1-0) and \ci\  data, and potential deep optical follow-up imaging, will allow us to investigate the physical properties of the ISM and CGM, and further reveal the nature of cold molecular medium.

\section{Acknowledgements}
Z.C. and J.L. are supported by the National Key R\&D Program of China (grant No. 2018YFA0404503), the National Science Foundation of China (grant No. 12073014) and Tsinghua University Initiative Scientific Research Program (grant No. 2019Z07L02017). We thank the anonymous referee for the very valuable feedback. This paper makes use of the following ALMA data: ADS/JAO.ALMA$\#$2019.1.01251.S. ALMA is a partnership of ESO (representing its member states), NSF (USA) and NINS (Japan), together with NRC (Canada), MOST and ASIAA (Taiwan), and KASI (Republic of Korea), in cooperation with the Republic of Chile. The Joint ALMA Observatory is operated by ESO, AUI/NRAO and NAOJ. The National Radio Astronomy Observatory is a facility of the National Science Foundation operated under cooperative agreement by Associated Universities, Inc. Based on observations made with the NASA/ESA Hubble Space Telescope, obtained from the data archive at the Space Telescope Science Institute, which is operated by the Association of Universities for Research in Astronomy, Incorporated, under NASA contract NAS5-26555. Support for Program number HST-GO-16891.002-A was provided through a grant from the STScI under NASA contract NAS5-26555.

\appendix

\section{Appendix: Total intensity, Mean velocity and Velocity dispersion map of Q0107}

\begin{figure*}
\centering
\includegraphics[width=1.0\textwidth]{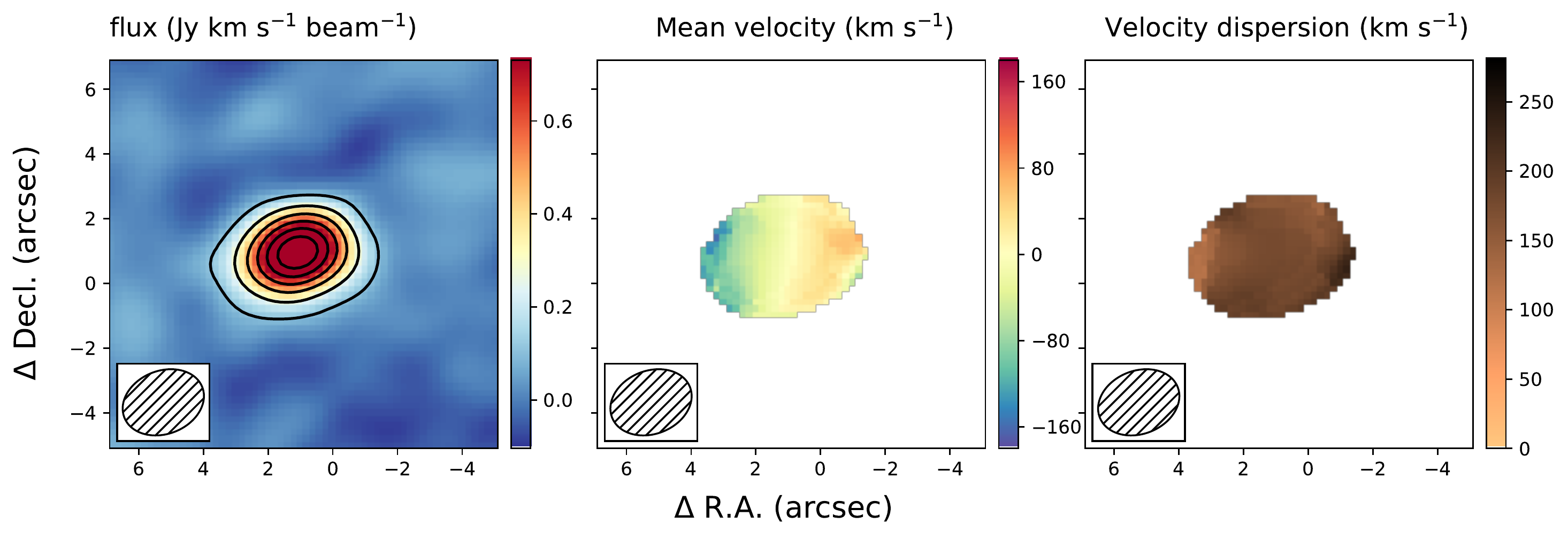}
\caption{The Total intensity  (left), Mean velocity  (middle) and Velocity dispersion  (right) maps of Q0107.} 
\label{fig:optical}
\end{figure*}

Fig.~\ref{fig:optical} shows the Total intensity, Mean velocity and Velocity dispersion map of the CO(4–3) emission from Q0107 within the velocity range: $(-482<v< 298)$ \kms. We derive these maps using Qubefit \footnote{https://github.com/mneeleman/qubefit} \citep{2020Neeleman}.

\section{Appendix: A Radial ($U$,$V$) Plot of Q2121}

\begin{figure}
\centering
\includegraphics[width=0.7\textwidth]{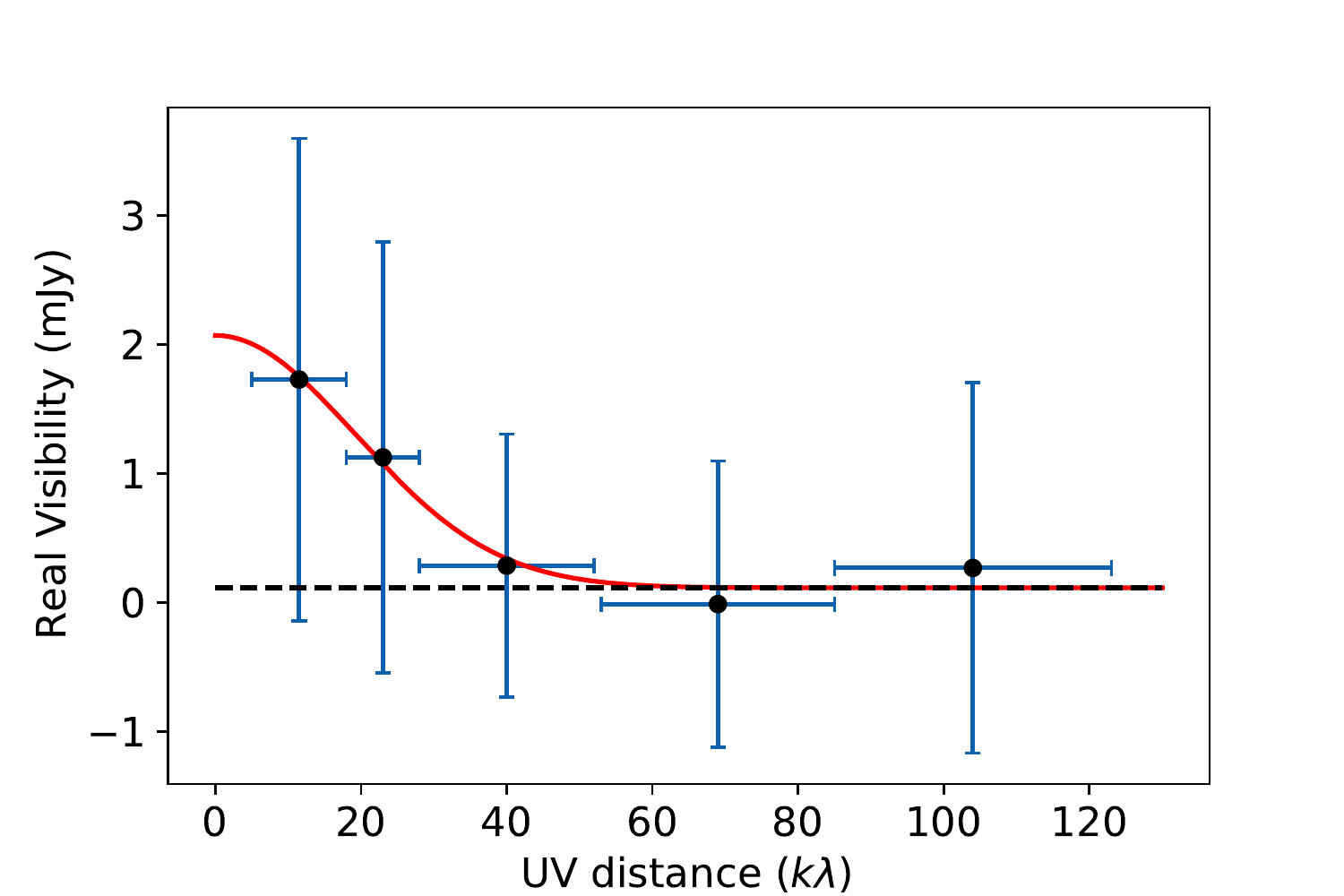}
\caption{($U$,$v$) plot of the 12m and 7m ACA combined CO(4–3) data of Q2121. Plotted are the real parts of the complex interferometer visibilities for the CO(4-3) signal per channel, when averaged across a velocity range of 240 \kms\ centered on the CO line and over all baselines within the ($u$,$v$)-range covered by the horizontal blue error bars. The vertical blue bars represent the error of the real part of the visibility amplitudes. The red line shows the fit of a model of a Gaussian distribution with a flux = $1.96 \pm 0.24$ mJy and a FWHM = ($4.0 \pm 0.2$)$^{\prime\prime}$. Across the 240 \kms\ that we used to average the ($u$,$v$) data, this corresponds to an integrated flux density of $0.47 \pm 0.06$ Jy \kms, which is consistent with $I_\nu$ in Table\,\ref{tab:gauss_fit}. The best-fit result is obtain by also including a CO point-source component with a flux density of 0.12 mJy, although the flux of this component is highly uncertain.} 
\label{fig:uv}
\end{figure}

Fig.~\ref{fig:uv} shows the ($u$,$v$) distance ($k \lambda$) plotted against the visibility amplitude of the CO(4–3) emission from Q2121. To derive this plot, we first split out the continuum-subtracted 12m and 7m ACA data sets to contain the velocity range ($-125 \sim 115$ \kms) respectively, and then combined them to extract all the data points. Finally, we binned these data points and fitted the Fourier transform of a Gaussian function and a constant function, which corresponds to the model of a Gaussian distribution and a point source in the image plane.

Among all of our spatially resolved sources, the reason why we only show the radial ($u$,$v$) analysis of Q2121 (which has no CO companions) is that, for the other sources, the CO emission from their close companions ($<$ 80 kpc) will be also included when we focus on the short baselines. This makes the results unreliable. In addition, a ($u$,$v$) analysis is more applicable to targets which extended emission shows an isotrophic halo structure, because for any ($u$,$v$) analysis, only emission that is distributed symmetrically around the phase center will be included at full amplitude. For our other five sources, because the extended gas is weaker and much more directional, ($u$,$v$) plots may not give significant results compared to what we obtained for Q1228 \citep{Li2021} and Q2121.

\bibliographystyle{aasjournal}

\end{document}